\documentclass[aps, prl, amsmath,amssymb,floatfix,superscriptaddress]{revtex4}
\usepackage{graphicx}
\usepackage{dcolumn}
\usepackage{bm}
\usepackage{hyperref} 
\usepackage{latexsym}
\usepackage{float}
\usepackage{rotating}

\def\df{d_{\text{f}}}
\def\tdf{\tilde d_{\text{f}}}

\def\dk{d_{\text{k}}}
\def\tdk{\tilde d_{\text{k}}}

\begin{document}
\title{Fractal and Transfractal Scale-Free Networks}   

\author{Hern\' an D. Rozenfeld}
\email{hernanrozenfeld@gmail.com}
\affiliation{Levich Institute and Physics Department, City College of New York, New York, New York 10031, USA}
\author{Lazaros K. Gallos}
\email{gallos@sci.ccny.cuny.edu}
\affiliation{Levich Institute and Physics Department, City College of New York, New York, New York 10031, USA}
\author{Chaoming Song}
\email{chaoming\_song@msn.com}
\affiliation{Levich Institute and Physics Department, City College of New York, New York, New York 10031, USA}
\author{Hern\'an A. Makse}
\email{hmakse@levdec.engr.ccny.cuny.edu}
\affiliation{Levich Institute and Physics Department, City College of New York, New York, New York 10031, USA}

\maketitle

\section*{Article Outline}
Glossary and Notation

\begin{enumerate}
\item Definition of the Subject and Its Importance
\item Introduction
\item Fractality in Real-World Networks
\item Models: Deterministic Fractal and Transfractal Networks
\item Properties of Fractal and Transfractal Networks
\item Future Directions
\item APPENDIX: The Box Covering Algorithms
\item Bibliography
\end{enumerate}

\section*{Glossary and Notation}

\noindent \textbf{Degree of a node} Number of edges incident to the node.
\\ \\
\textbf{Scale-Free Network} Network that exhibits a wide (usually power-law) distribution of the degrees.
\\ \\
\textbf{Small-World Network} Network for which the diameter increases logarithmically with the number of nodes.
\\ \\
\textbf{Distance} The length (measured in number of links) of the shortest path between two nodes.
\\ \\
\textbf{Box} Group of nodes. In a \emph{connected box} there exists a path within the box between any pair of nodes. Otherwise, the box is \emph{disconnected}.
\\ \\
\textbf{Box Diameter} The longest distance in a box.
\\ \\

\section{{\large Definition of the Subject and Its Importance}}

The explosion in the study of complex networks during the last decade has offered a unique view in
the structure and behavior of a wide range of systems, spanning many different disciplines~\cite{reviews}.
The importance of complex networks lies mainly in their simplicity, since they can represent practically
any system with interactions in a unified way by stripping complicated details and retaining
the main features of the system. The resulting networks include only \emph{nodes}, representing the interacting agents
and \emph{links}, representing interactions.
The term `interactions' is used loosely to describe any possible way that causes two nodes
to form a link. Examples can be real physical links, such as the wires connecting computers in the
Internet or roads connecting cities, or alternatively they may be virtual links, such as links in WWW homepages
or acquaintances in societies, where there is no physical medium actually connecting the nodes.

The field was pioneered by the famous mathematician P. Erd\H{o}s many decades ago, when he greatly advanced graph theory~\cite{erdos}.
The theory of networks would have perhaps remained a problem of mathematical beauty, if it was not for the discovery that
a huge number of everyday life systems share many common features and can thus be described through a unified theory.
The remarkable diversity of these systems incorporates artificially or man-made technological
networks such as the Internet and the World Wide Web (WWW), social networks such as social acquaintances or
sexual contacts, biological networks of natural origin, such as the network of protein
interactions of Yeast~\cite{reviews, vidal2}, and a rich variety of other systems, such as proximity of words in literature~\cite{motter}, items that are bought by the same people~\cite{butler} or the way modules are connected to create a piece of software, among many others.

The advances in our understanding of networks, combined with the increasing availability of many databases,
allows us to analyze and gain deeper insight into the main characteristics of these complex systems.
A large number of complex networks share the \emph{scale-free} property~\cite{reviews, faloutsos1}, indicating the presence of few highly connected nodes
(usually called hubs) and a large number of nodes with small degree. This feature alone has a great impact on the analysis of complex networks and has
introduced a new way of understanding these systems.
This property carries important implications in many everyday life problems, such as the way a disease spreads in communities of individuals, or the resilience and
tolerance of networks under random and intentional attacks~\cite{cohen1, cohen2, cohen3, schwartz, lazaros1}.

Although the scale-free property holds an undisputed importance, it has been shown to not completely determine the global structure of networks~\cite{jmatrix}.
In fact, two networks that obey the same distribution of the degrees may dramatically differ in other fundamental structural properties, such as in correlations
between degrees or in the average distance between nodes.
Another fundamental property, which is the focus of this article, is the presence
of self-similarity or fractality. In simpler terms, we want to know
whether a subsection of the network looks much the same as the whole~\cite{bunde-havlin, vicsek, feder, bahbook}.
In spite of the fact that in regular fractal objects the distinction between self-similarity and fractality is absent, in network theory we can
distinguish the two terms: in a \emph{fractal network} the number of boxes of a given size that are needed to completely cover the network
scales with the box size as a power law,
while a \emph{self-similar network} is defined as a network whose degree distribution remains invariant under renormalization of the network (details 
on the renormalization process will be provided later).
This essential result allows us to better understand the origin of important structural properties of networks such as the power-law degree distribution~\cite{song, song2, kim0}.

\section{Introduction}

Self-similarity is a property of fractal structures, a concept
introduced by Mandelbrot and one of the fundamental mathematical
results of the 20th century~\cite{mandelbrot,vicsek,feder}.  The
importance of fractal geometry stems from the fact that these
structures were recognized in numerous examples in Nature, from the
coexistence of liquid/gas at the critical point of evaporation of
water~\cite{kadanoff2,stanley,cp}, to snowflakes, to the tortuous
coastline of the Norwegian fjords, to the behavior of many
complex systems such as economic data, or the complex patterns
of human agglomeration~\cite{vicsek,feder}.

Typically, real world scale-free networks exhibit the small-world property~\cite{reviews}, which implies that the number of nodes increases exponentially with the diameter of the network, rather than the power-law behavior expected for self-similar structures. For this reason complex networks were believed to \emph{not} be length-scale invariant or self-similar. 

In 2005, C. Song, S. Havlin and H. Makse presented an approach to analyze complex networks, that reveals their self-similarity~\cite{song}. This result is achieved by the application of a renormalization procedure which coarse-grains the system into boxes containing nodes within a given size~\cite{song, song3}. As a result, a power-law relation between the number of boxes needed to cover the network and the size of the box is found, defining a finite self-similar exponent.
These fundamental properties, which are shown for the WWW, cellular and  protein-protein interaction networks, help to understand the emergence of the scale-free property in complex networks. They suggest a common self-organization dynamics of diverse networks at different scales into a critical state and in turn bring together previously unrelated fields: the statistical physics of complex networks with renormalization group, fractals and critical phenomena.

\section{Fractality in Real-World Networks}

The study of real complex networks has revealed that many of them share some fundamental
common properties. Of great importance is the form of the degree distribution for
these networks, which is unexpectedly wide. This means that the degree of a node
may assume values that span many decades. Thus, although the majority of nodes have a relatively
small degree, there is a finite probability that a few nodes
will have degree of the order of thousands or even millions. Networks that exhibit such a wide distribution
$P(k)$ are known as \emph{scale-free} networks, where the term refers to the absence of a characteristic
scale in the degree $k$. This distribution very often obeys a power-law form with a degree exponent
$\gamma$, usually in the range $2<\gamma<4$ \cite{barabasi1999},
\begin{equation}
P(k) \sim k^{-\gamma} \,.
\label{scale-free}
\end{equation}

A more generic property, that is usually inherent in scale-free networks but applies equally
well to other types of networks, such as in Erd\H{o}s-R\'enyi random graphs, is the \emph{small-world}
feature. Originally discovered in sociological studies~\cite{milgram}, it is the generalization of
the famous `six degrees of separation' and refers to the very small network diameter.
Indeed, in small-world networks a very small number of steps is required to reach a given
node starting from any other node. Mathematically this is expressed by the slow (logarithmic) increase
of the average diameter of the network, $\bar{\ell}$, with the
total number of nodes $N$, $\bar{\ell} \sim \ln N$, where $\ell$
is the {\it shortest} distance between two nodes and defines the
distance metric in complex networks
\cite{erdos,bollobas,watts,barabasi1999}, namely,
\begin{equation}
N \sim e^{\bar{\ell}/\ell_0}, \label{smallworld}
\end{equation}
where $\ell_0$ is a characteristic length.

\begin{figure}
\centering{ { \resizebox{10cm}{!} {\includegraphics{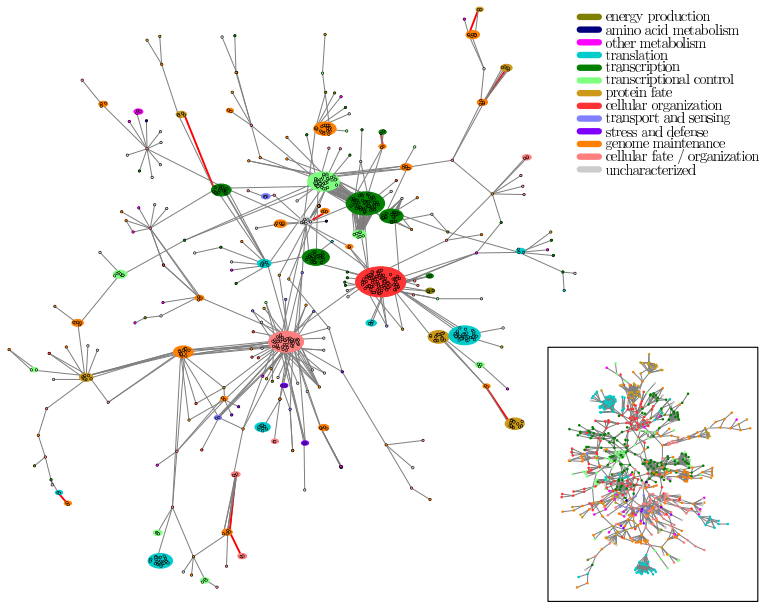}}}
}
\caption{\label{pin} Representation of the Protein Interaction Network of Yeast. The colors show different subgroups of proteins that participate in different functionality classes~\cite{vidal2}.}
\end{figure}

These network characteristics have been shown to apply in many empirical studies of diverse
systems \cite{reviews, faloutsos1,barabasi1999}. The simple knowledge
that a network has the scale-free and/or small-world property already enables us to qualitatively
recognize many of its basic properties. However, structures that have the same degree exponents
may still differ in other aspects~\cite{jmatrix}. For example, a question of fundamental importance is
whether scale-free networks are also self-similar or fractals. The illustrations of scale-free
networks (see e.g. Figs.~\ref{pin} and~\ref{web}b) seem to resemble traditional fractal objects.
Despite this similarity, Eq.~(\ref{smallworld}) definitely appears to contradict
a basic property of fractality: fast increase of the diameter with the system size.
Moreover, a fractal object should be self-similar or invariant under a scale transformation,
which is again not clear in the case of scale-free networks where the scale has
necessarily limited range. So, how is it even possible that fractal scale-free networks exist?
In the following, we will see how these seemingly contradictory aspects can be reconciled.

\subsection{Fractality and Self-Similarity}

The classical theory of self-similarity requires a power-law relation
between the number of nodes $N$ and the diameter of a fractal object $\ell$~\cite{bunde-havlin,bahbook}.
The fractal dimension can be calculated using either \emph{box-counting} or \emph{cluster-growing}
techniques \cite{vicsek}. In the first method the network is covered with $N_B$ boxes of
linear size $\ell_B$. The fractal dimension or box dimension $d_B$
is then given by \cite{feder}:
\begin{equation}
N_B \sim \ell_B^{-d_B} \,.
\label{dh}
\end{equation}
In the second method, instead of covering the network with boxes, a random
seed node is chosen and nodes centered at the seed are grown so that they are
separated by a maximum distance $\ell$.
The procedure is then repeated by choosing many seed nodes
at random and the average ``mass'' of the resulting
clusters, $\langle M_c\rangle$ (defined as the number of nodes in the cluster)
is calculated as a function of $\ell$
to obtain the following scaling:
\begin{equation}
\langle M_c\rangle  \sim \ell^{d_f},
\label{df}
\end{equation}
defining the fractal cluster dimension $d_f$ \cite{feder}.
If we use Eq. (\ref{df}) for a small-world network, then Eq.~(\ref{smallworld}) readily
implies that $d_f=\infty$. In other words, these networks cannot be
characterized by a finite fractal dimension, and should be regarded as infinite-dimensional objects. If this were true, though, local properties in a part of the network would not be able to represent the
whole system. Still, it is also well established that the scale-free nature is similar
in different parts of the network. Moreover, a graphical representation of real-world networks allows us to see that those systems seem to be built by attaching (following some rule) copies of itself.

The answer lies in the inherent inhomogeneity of the network. In the classical
case of a \emph{homogeneous} system (such as a fractal percolation cluster) the degree
distribution is very narrow and the two methods described above are fully equivalent,
because of this local neighborhood invariance.
Indeed, all boxes in the box-covering method are statistically similar with each other
as well as with the boxes grown when using the cluster-growing technique, so that 
Eq.~(\ref{df}) can be derived from Eq.~(\ref{dh}) and $d_B=d_f$.

\begin{figure}
\centerline{
{\bf a}{ \resizebox{10cm}{!}{\includegraphics{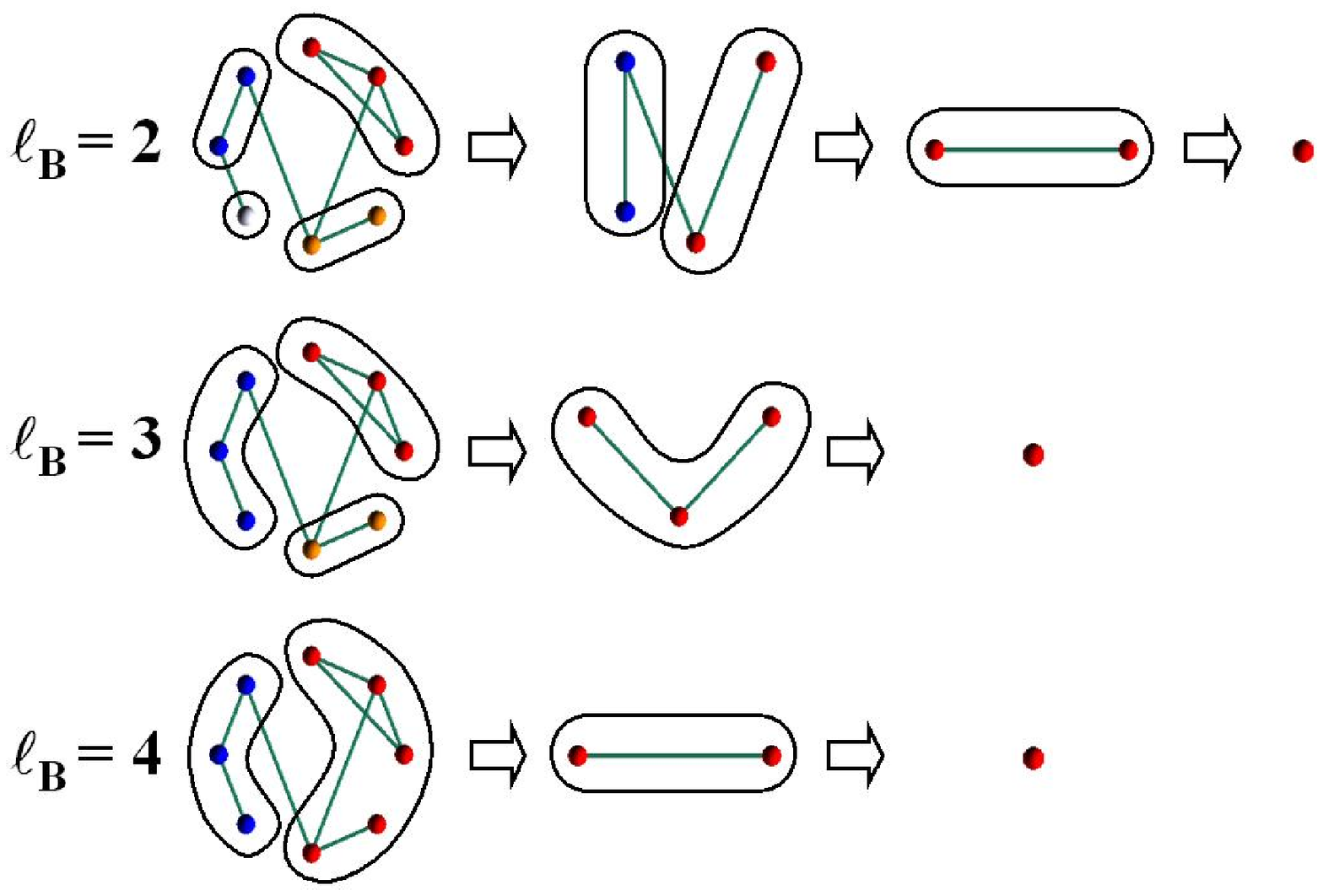}}}}
\centerline{
{\bf b}
{ \resizebox{10cm}{!}{\includegraphics{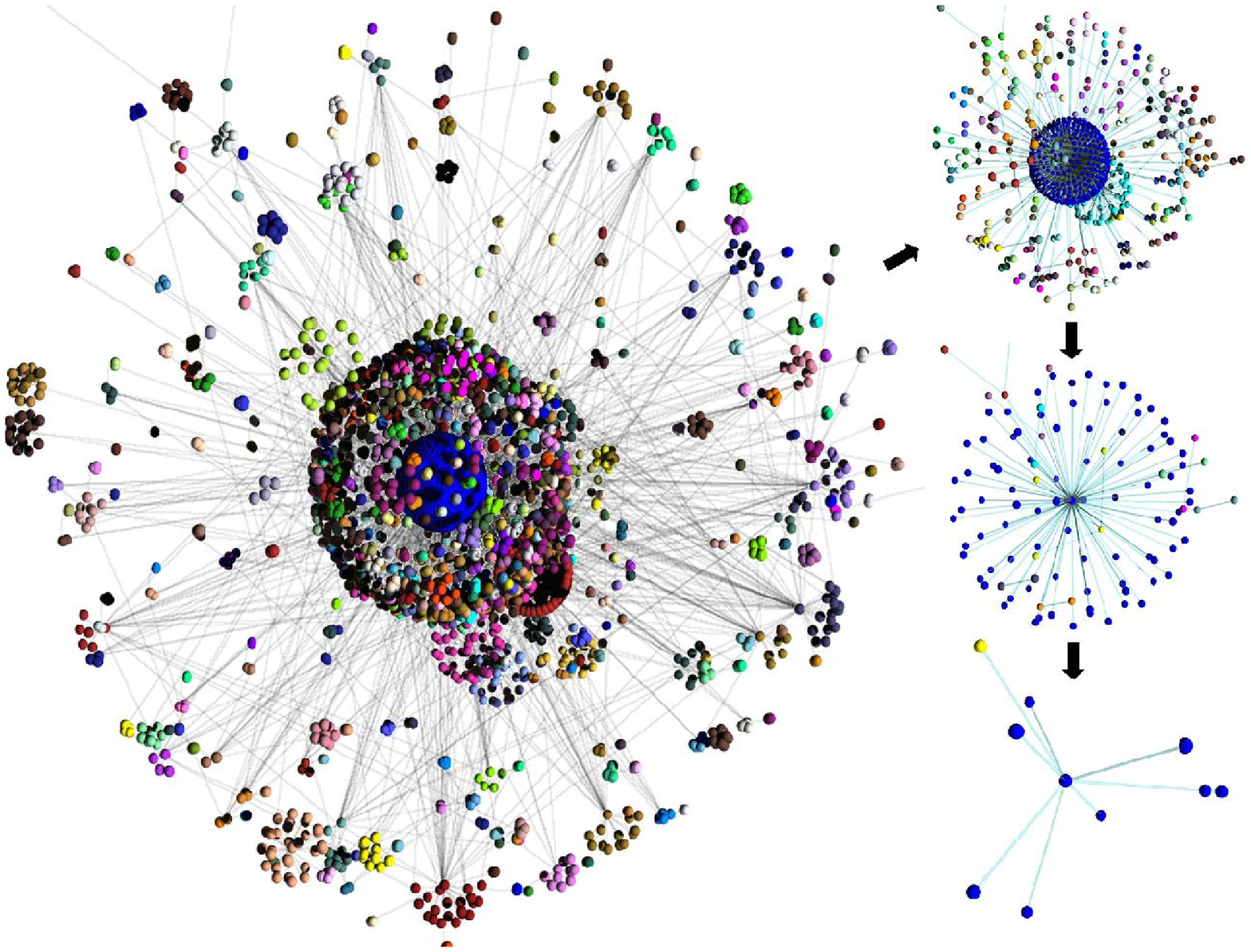}}}
}
\caption{The renormalization procedure for complex networks.
\bf a,} Demonstration of the method
for different $\ell_B$ and different stages in a network demo.
The first column depicts the original network.
The system is tiled with boxes of size $\ell_B$
(different colors correspond to different boxes).
 All nodes in
a box are connected by a minimum distance smaller than the given $\ell_B$.
For instance, in the case of
$\ell_B=2$, one identifies four boxes which contain the nodes depicted
with colors red, orange, white, and blue, each containing 3, 2, 1, and
2 nodes, respectively.
Then each box is replace by a single node; two renormalized nodes are
connected if there is at least one link
between the unrenormalized boxes.
 Thus we obtain the network
shown in the second column.
The resulting number of boxes needed to tile
the network, $N_B(\ell_B)$,
is plotted in Fig. \protect\ref{fractal}
versus $\ell_B$ to
obtain   $d_B$
as in Eq. (\protect\ref{dh}).
The renormalization procedure is applied again
and repeated until
the network is reduced to a single node
(third and fourth columns for different $\ell_B$).
{\bf b,} Three stages in the renormalization scheme applied to the entire WWW.
We fix the box size to  $\ell_B = 3$ and apply the
renormalization for four stages. This corresponds, for instance,
to the sequence for the network demo
depicted in the second row in part {\bf a} of this figure.
We color the nodes in the web according to the boxes to which
they belong.
\label{web}
\end{figure}
In \emph{inhomogeneous} systems, though, the local environment can vary significantly.
In this case, Eqs. (\ref{dh}) and (\ref{df}) are no longer equivalent. If we focus on
the box-covering technique then we want to cover the entire network with the minimum
possible number of boxes $N_B(\ell_B)$, where the distance between any two nodes that belong in
a box is smaller than $\ell_B$. An example is shown in Fig.~\ref{web}a using a simple
8-node network. After we repeat this procedure for different values of $\ell_B$
we can plot $N_B$ vs $\ell_B$. 

\begin{figure}
\centering
\hbox{
{\bf a}{ \resizebox{7.9cm}{!}{\includegraphics{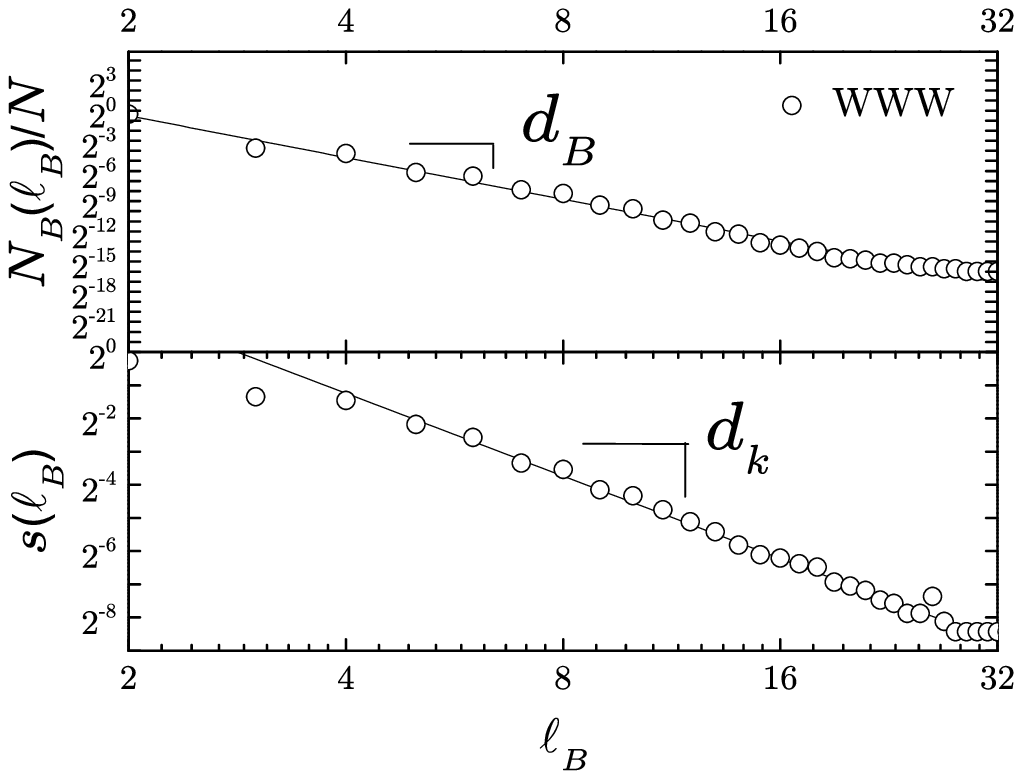}}}

\centering
{\bf b}{ \resizebox{8cm}{!}{\includegraphics{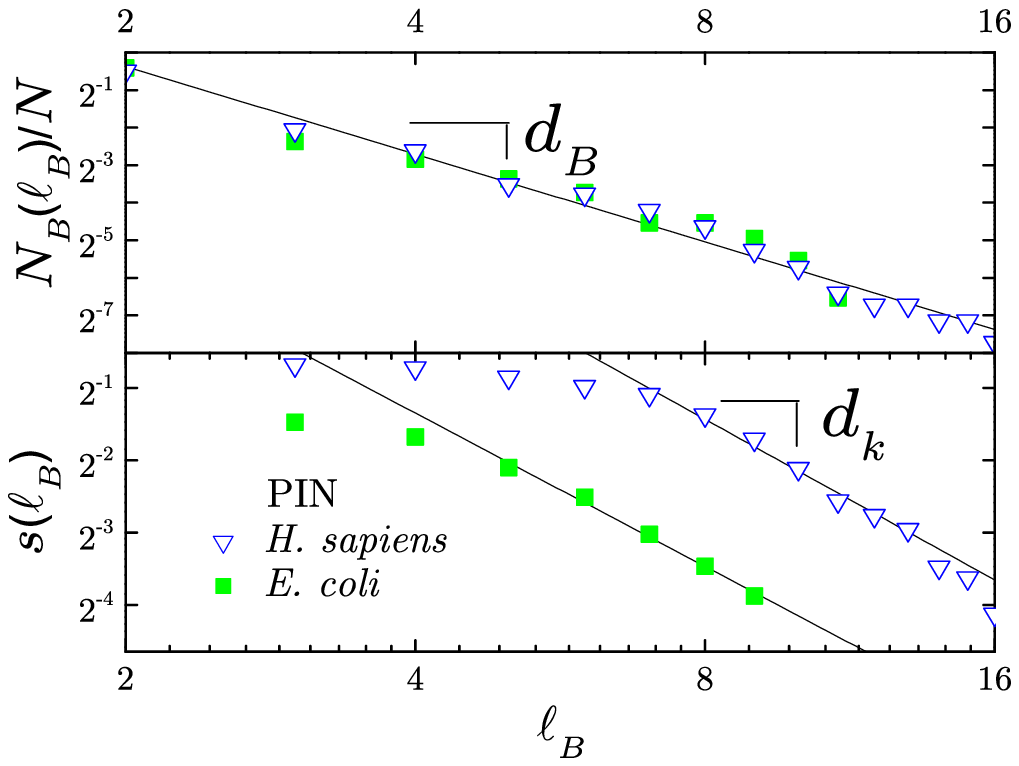}}}
}
\centering
{\bf c}{ \resizebox{8cm}{!}{\includegraphics{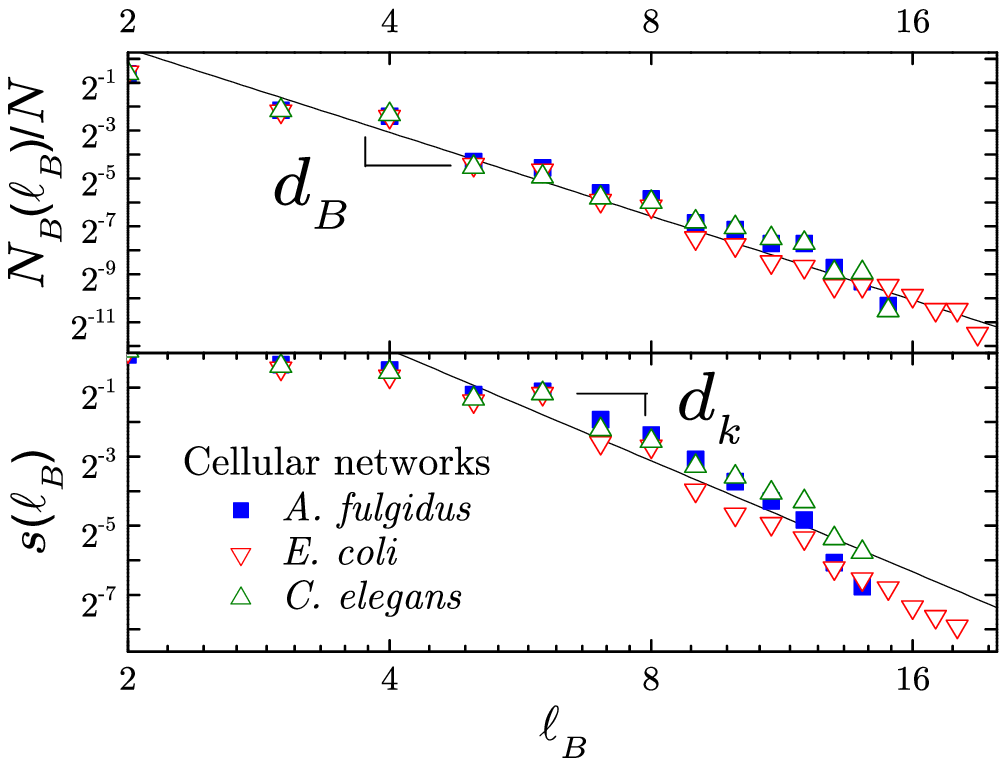}}}
\centering
\\
{\bf d}{ \resizebox{8cm}{!}{\includegraphics{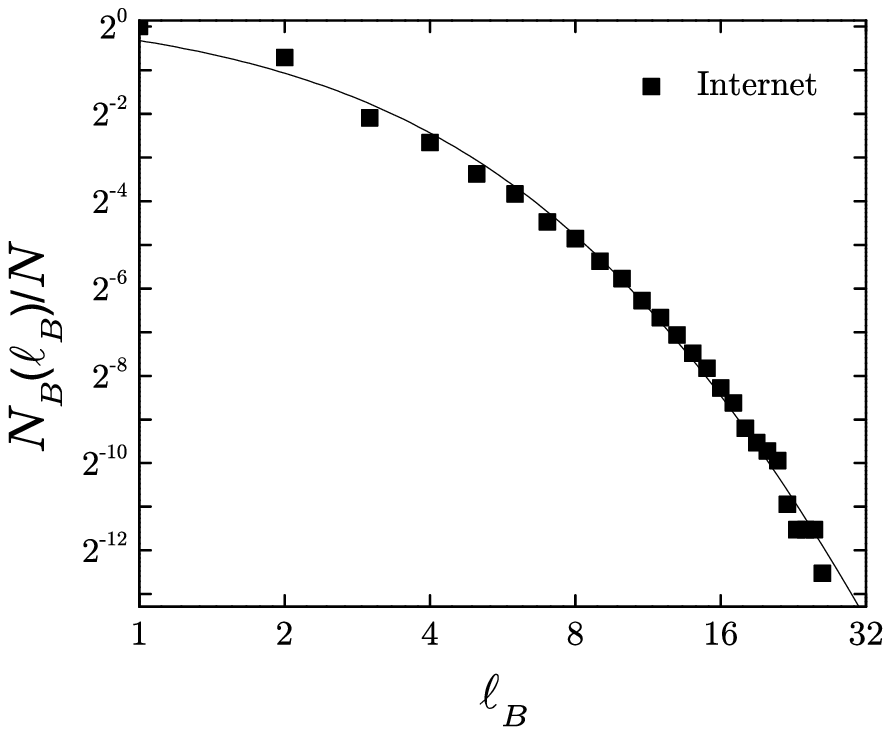}}}\centering
{\bf e}{ \resizebox{8cm}{!}{\includegraphics{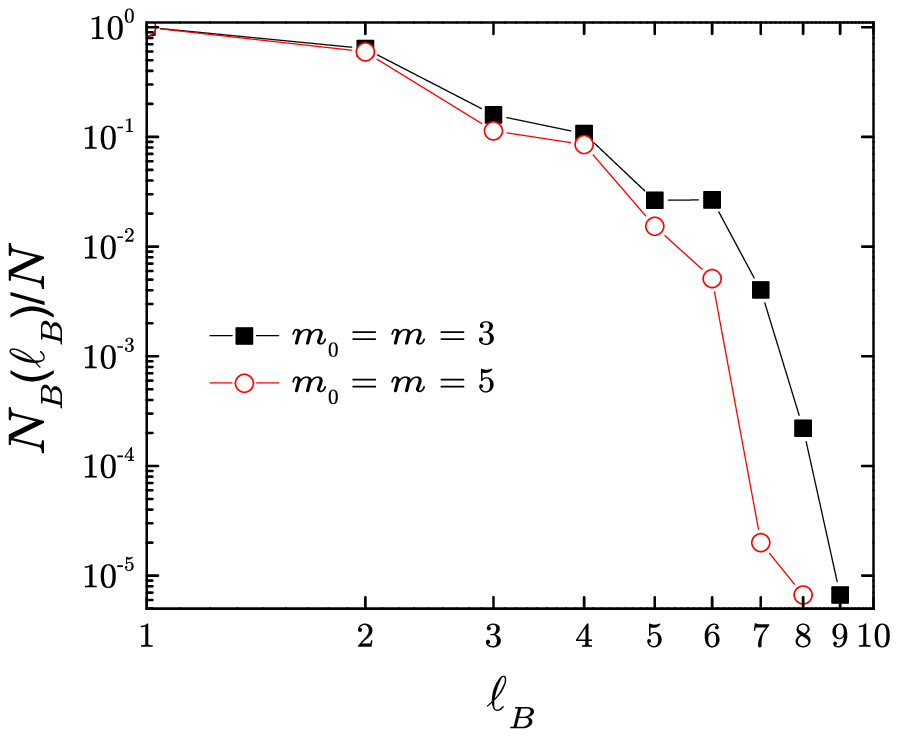}}}
\caption{Self-similar scaling in complex networks.
{\bf a,}  Upper panel: Log-log plot of the $N_B$ vs $\ell_B$ revealing
the
self-similarity of the WWW according to Eq. (\ref{dh}). Lower panel: The scaling of $s(\ell_B)$
vs. $\ell_B$ according to Eq. (\ref{sl}).
{\bf b,} Same as (a) but for two
protein interaction networks: {\it H. sapiens} and {\it E. coli}.
Results are analogous to (b) but with different scaling exponents.
{\bf c,} Same as (a) for the cellular networks of
{\it A. fulgidus}, {\it E. coli} and {\it C. elegans}.
{\bf d,} Internet. Log-log plot of $N_B(\ell_B)$. The solid line shows that the internet~\cite{mbone} is \emph{not} a fractal network since in does not follow the power-law relation of Eq.(\ref{mass_b}).
{\bf e,} Same as (d) for the Barab\'asi-Albert model network~\cite{barabasi2} with $m=3$ and $m=5$.
}
\label{scaling}
\end{figure}

When the box-covering method is applied to real large-scale networks, such us the WWW~\cite{barabasi1999} (http://www.nd.edu/$\sim$networks), the network of protein interaction of \emph{H. sapiens} and \emph{E. coli}~\cite{xenarios,PIN} and several cellular networks~\cite{barabasi1,overbeek}, then they follow Eq.~(\ref{dh}) with a clear power-law, indicating the \emph{fractal} nature of these systems (Figs.~\ref{scaling}a,~\ref{scaling}b,~\ref{scaling}c). On the other hand when the method is applied to other real world networks such us the Internet~\cite{mbone} or the Barab\'asi-Albert network~\cite{barabasi2}, they do not satisfy Eq.~(\ref{dh}), which manifests that these networks are \emph{not} fractal.

The reason behind the discrepancy in the fractality of homogeneous and inhomogeneous systems can be better clarified studying the mass of the
boxes. For a given $\ell_B$ value, the average mass of a box $\langle M_B (\ell_B)\rangle$ is
\begin{equation}
\langle M_B (\ell_B)\rangle  \equiv N / N_B(\ell_B) \sim \ell_B^{d_B} \,,
\label{mass_b}
\end{equation}
as also verified in Fig.~\ref{scaling} for several real-world networks. On the other hand, the average performed in the cluster growing
method (averaging over single boxes without tiling the system)
gives rise to an exponential growth of the mass
\begin{equation}
\langle M_c(\ell)\rangle  \sim e^{\ell/\ell_1},
\label{mass_c}
\end{equation}
in accordance with the small-world effect, Eq. (\ref{smallworld}).
Correspondingly, the probability distribution of the mass of the boxes $M_B$ using box-covering is very broad, while the cluster-growing technique leads to a narrow probability distribution of
$M_c$.


The topology of scale-free networks is dominated by several highly
connected hubs--- the nodes with the largest degree--- implying
that most of the nodes are connected to the hubs via one or very
few steps. Therefore, the average performed in the cluster growing
method is biased; the hubs are overrepresented in Eq. (\ref{mass_c})
since almost every node is a neighbor of a hub, and there is always a very large
probability of including the same hubs in all clusters.
On the other hand, the box covering method is a global tiling of the system
providing a flat average over all the nodes, i.e. each part of the
network is covered with an equal probability. Once a hub (or any
node) is covered, it cannot be covered again.

In conclusion, we can state that the two dominant methods that are routinely
used for calculations of fractality and give rise to Eqs.~(\ref{dh}) and (\ref{df})
are not equivalent in scale-free networks, but rather highlight different aspects:
box covering reveals the self-similarity, while cluster growth reveals the small-world
effect. The apparent contradiction is due to the hubs being used many times
in the latter method.

The apparent contradiction between the small-world and fractal properties, as expressed through
Eqs.~\ref{smallworld} and \ref{dh} can be explained as follows.
Scale-free networks can be classified into three groups:
(i) pure fractal, (ii) pure small-world and (iii) a mixture between fractal and small-world. (i) A fractal network
satisfies Eq.~\ref{dh} at all scales, meaning that for any value of $\ell_B$, the number of boxes always follows
a power-law (examples are shown in Fig.~\ref{scaling}a,~\ref{scaling}b,~\ref{scaling}c). (ii) When a network is a
pure small-world, it never satisfies Eq.~\ref{dh}. Instead, $N_B$ follows an exponential decay with $\ell_B$ and
the network cannot be regarded as fractal. Figs.~\ref{scaling}d and~\ref{scaling}e show two examples of pure small-world
networks. (iii) In the case of mixture between fractal and small-world, Eq.~\ref{dh} is satisfied up to
some cut-off value of $\ell_B$, above which the fractality breaks down and the small-world property emerges. The small-world
property is reflected in the plot of $N_B$ vs. $\ell_B$ as an exponential cut-off for large $\ell_B$.

We can also understand the coexistence of the small-world property and the fractality through a more intuitive approach. In a pure
fractal network the length of a path between any pair of nodes scales as a power-law with the number of nodes in the network. Therefore,
the diameter $L$ also follows a power-law, $L \sim N^{1/d_B}$. If one adds a few shortcuts (links between randomly chosen nodes), many
paths in the network are drastically shortened and the small-world property emerges as $L \sim \textrm{Log}~N$. In spite of this fact,
for shorter scales, $\ell_B \ll L$, the network still behaves as a fractal. In this sense, we can say that globally the network is small-world,
but locally (for short scales) the network behaves as a fractal. As more shortcuts are added, the cut-off in a plot of $N_B$ vs. $\ell_B$
appears for smaller $\ell_B$, until the network becomes a pure small-world for which all paths lengths increase logarithmically with $N$.

The reasons why certain networks have evolved towards a fractal or non-fractal
structure will be described later, together with models and examples that provide additional
insight into the processes involved.

\subsection{Renormalization}

Renormalization is one of the most important techniques in modern Statistical Physics~\cite{cardy, kadanoff,salmhofer}.
The idea behind this procedure is to continuously create smaller replicas of a given
object, retaining at the same time the essential structural features, and hoping
that the coarse-grained copies will be more amenable to analytic treatment.

The idea for renormalizing the network emerges naturally from the concept of fractality described above. If a network is self-similar, then it will look more or less
the same under different scales. The way to observe these different length-scales
is based on renormalization principles, while the criterion to decide on whether
a renormalized structure retains its form is the invariance of the main structural
features, expressed mainly through the degree distribution.

The method works as follows. Start by fixing the value of $\ell_B$ and
applying the box-covering algorithm in order to cover the entire network
with boxes (see Appendix). In the renormalized network each box is replaced by a single
node and two nodes are connected if there existed at least one connection
between the two corresponding boxes in the original network. The resulting
structure represents the first stage of the renormalized network. We can
apply the same procedure to this new network, as well, resulting in the
second renormalization stage network, and so on until we are left with a single
node.

The second column of the panels in Fig. \ref{web}a shows this step
in the renormalization procedure for the schematic network, while Fig.
\ref{web}b shows the results for the same procedure applied to the
entire WWW for $\ell_B=3$.

\begin{figure}
\centering{ { \resizebox{8.8cm}{!} {\includegraphics{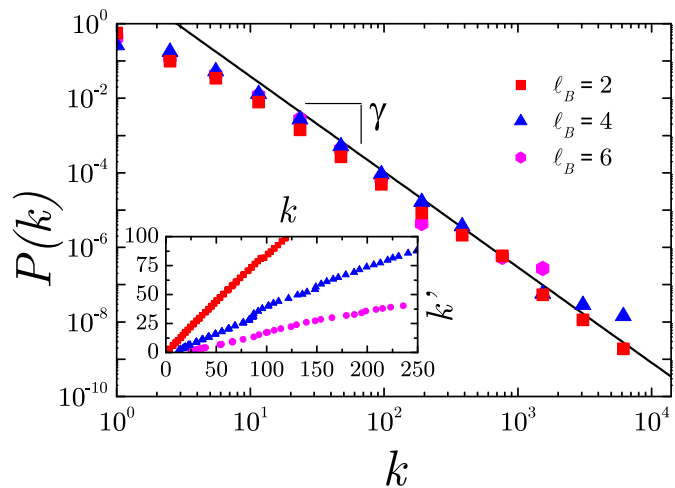}}}
}
\caption{\label{fractal}  Invariance of 
the degree distribution of the WWW under the renormalization for different box sizes, lB . 
We show the data collapse of the degree distributions demonstrating the self-similarity at 
different scales. The inset shows the scaling of $k'= s(\ell_{B})k$ for different $\ell_B$ , from where we 
obtain the scaling factor $s(\ell_{B})$. Moreover, renormalization for a fixed box 
size ($\ell_B = 3$) is applied, until the network is reduced to 
a few nodes. It was found that $P(k)$ is invariant under these multiple renormalizations procedures.}
\end{figure}

The renormalized network gives rise to a new probability
distribution of links, $P(k')$ (we use a prime $'$ to denote quantities
in the renormalized network). This distribution remains invariant
under the renormalization:
\begin{equation}
P(k)\rightarrow P(k') \sim (k')^{-\gamma}.
\label{pk}
\end{equation}
Fig.~\ref{fractal} supports the validity of this scale
transformation by showing a data collapse of all distributions
with the same $\gamma$ according to (\ref{pk}) for the WWW.

Here, we present the basic scaling relations that characterize renormalizable
networks. The degree $k'$ of each node in the renormalized network
can be seen to scale with the largest degree $k$ in the corresponding original box as
\begin{equation}
k \rightarrow k' = s(\ell_B) k \,.
\label{s}
\end{equation}
This equation defines the scaling transformation in the
connectivity distribution. Empirically, it was found that the
scaling factor $s(<1)$ scales with $\ell_B$ with a new exponent,
$d_k$, as $s(\ell_B) \sim \ell_B^{-d_k}$, so that
\begin{equation}
k' \sim \ell_B^{-d_k} k,
\label{sl}
\end{equation}
This scaling is verified for many networks, as shown in Fig. \ref{fractal}a.

The exponents $\gamma$, $d_B$, and $d_k$ are not all independent from
each other. The proof starts from the density balance equation $n(k) dk = n'(k') dk'$, where
$n(k) = N P(k)$ is the number of nodes with degree $k$ and $n'(k')
= N' P(k')$ is the number of nodes with degree $k'$ after the
renormalization ($N'$ is the total number of nodes in the
renormalized network). Substituting Eq. (\ref{s}) leads to $N' = s^{\gamma-1} N$.
Since the total number of nodes in the renormalized network is
the number of boxes needed to cover the unrenormalized network at
any given $\ell_B$ we have the identity $N' =N_B(\ell_B)$. Finally, from Eqs.
(\ref{dh}) and (\ref{sl}) one obtains the relation between the
three indexes
\begin{equation}
\gamma = 1 + d_{B} / d_{k}.
\label{relation}
\end{equation}

The use of Eq. (\ref{relation}) yields the same $\gamma$ exponent as
that obtained in the direct calculation of the degree
distribution. The significance of this result is that the
scale-free properties characterized by $\gamma$ can be related to
a more fundamental length-scale invariant property,
characterized by the two new indexes $d_{B}$ and $d_{k}$.

We have seen, thus, that concepts
introduced originally for the study of critical phenomena in
statistical physics, are also valid in the
characterization of a different class of phenomena: the topology
of complex networks. A large number of scale-free networks are fractals
and an even larger number remain invariant under a scale-transformation.
The influence of these features on the network properties will be delayed
until the sixth chapter, after we introduce some algorithms for efficient
numerical calculations and two theoretical models that give rise to fractal networks.

\section{Models: Deterministic Fractal and Transfractal Networks}

The first model of a scale-free fractal network was presented in 1979 when N. Berker and S. Ostlund~\cite{berker1} proposed a hierarchical network that served as an exotic example where renormalization group techniques yield exact results, including the percolation phase transition and the $q \rightarrow 1$ limit of the Potts model. Unfortunately, in those days the importance of the power-law degree distribution and the concept of fractal and non-fractal complex networks were not known. Much work has been done on these types of hierarchical networks. For example, in 1984, M. Kaufman and R. Griffiths made use of Berker and Ostlund's model to study the percolation phase transition and its percolation exponents~\cite{hierarchical1,hierarchical2,comellas}. 

Since the late 90's, when the importance of the power-law degree distribution was first shown~\cite{reviews} and after the finding of C. Song, S. Havlin and H. Makse~\cite{song}, many hierarchical networks that describe fractality in complex networks have been proposed. These artificial models are of great importance since they provide insight into the origins and fundamental properties that give rise to the fractality and non-fractality of networks. 

\subsection{The Song-Havlin-Makse Model}

\begin{figure*}
\centerline{ \hbox { (a)\resizebox{6cm}{!} {
\includegraphics{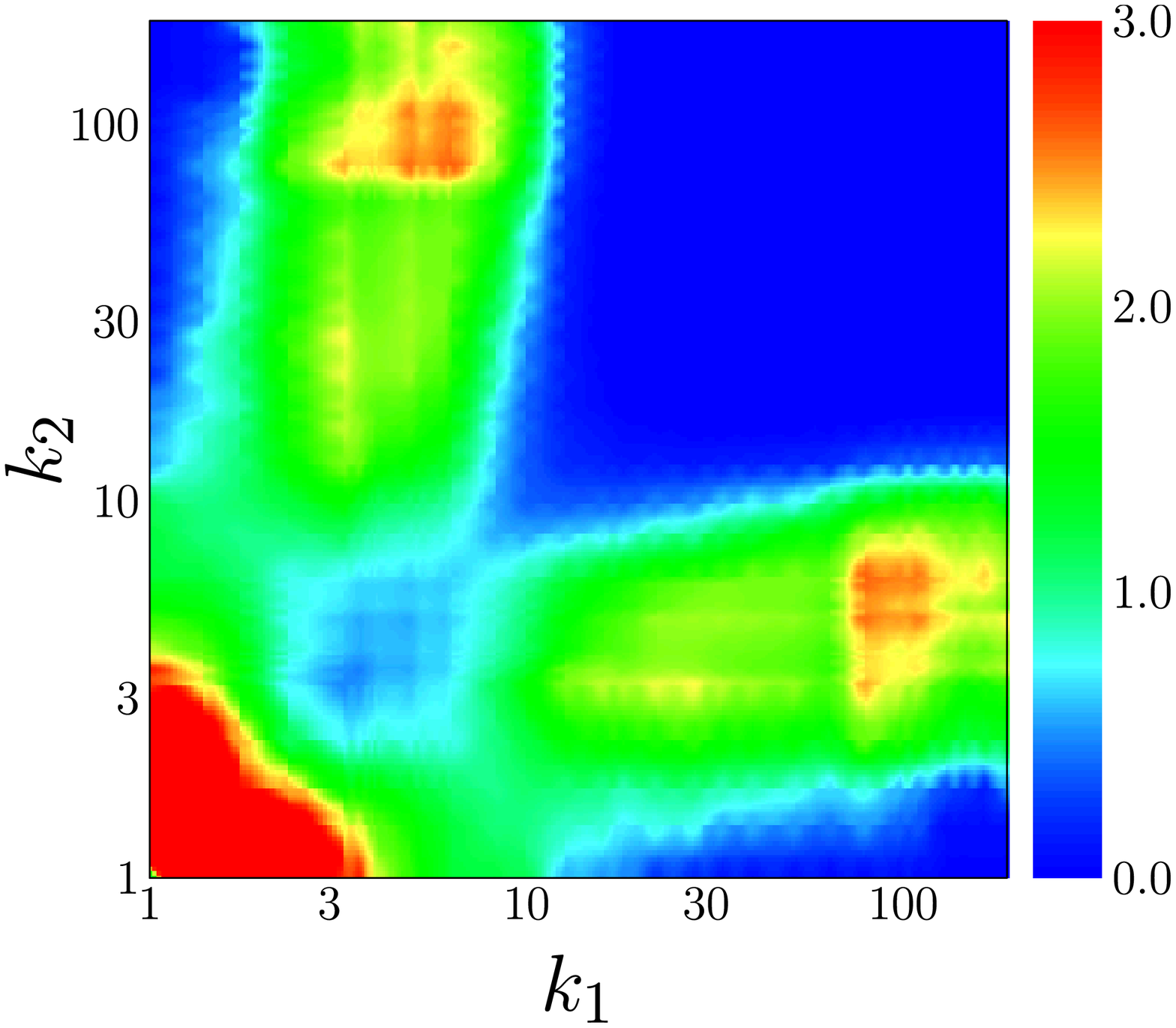}}
(b)\resizebox{6cm}{!} { \includegraphics{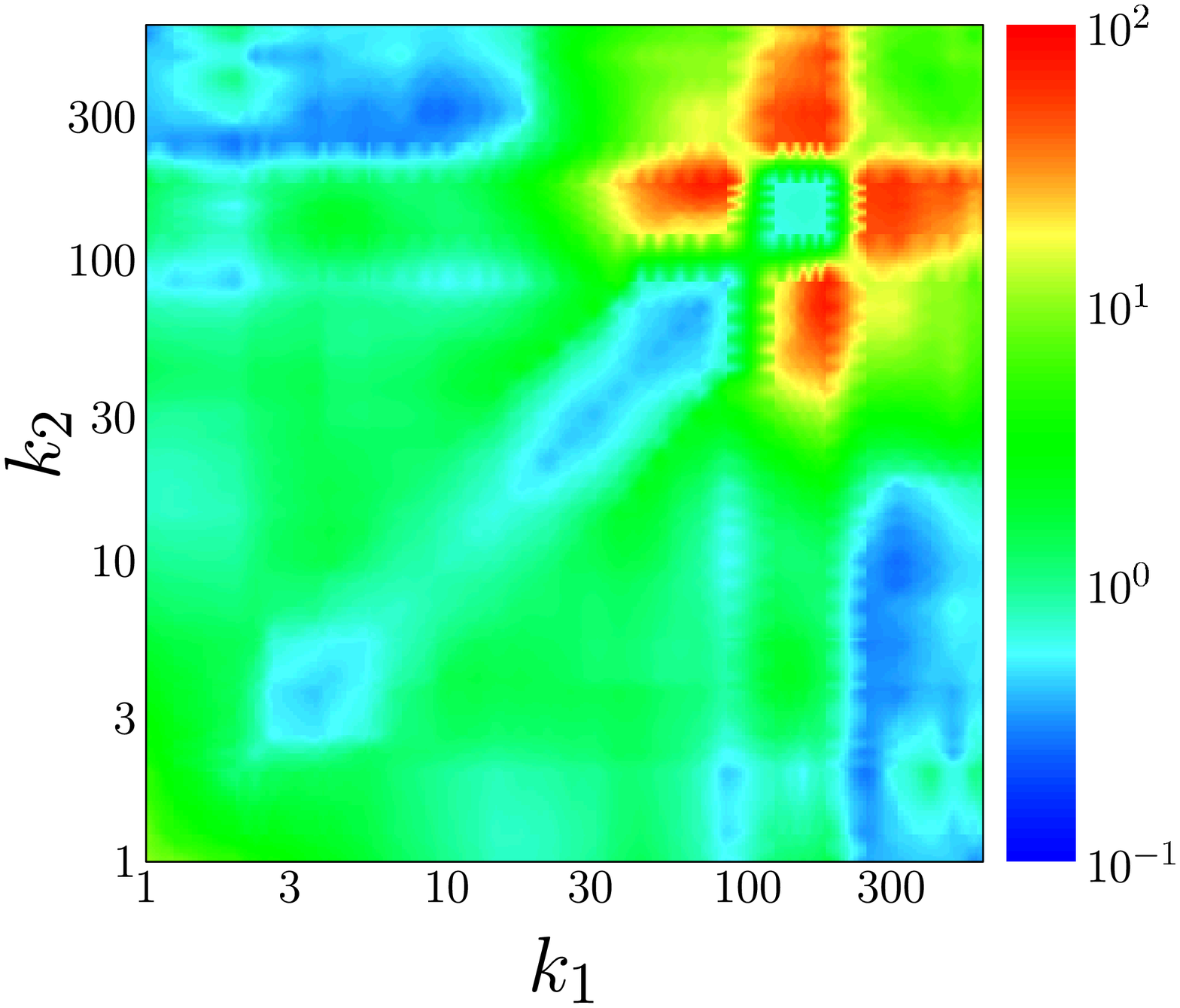}} }}
\caption {Degree correlation profile for (a) the cellular metabolic network of E. \emph{coli}, and (b) the Internet at the router level.} 
\label{kk}
\end{figure*}

The correlations between degrees in a network~\cite{newman1,newman2,maslov1,pastor1} are quantified through
the probability $P(k_{1},k_{2})$ that a node of degree $k_{1}$ is connected to another node of degree $k_{2}$. In Fig.~\ref{kk} we can see the degree correlation profile $R(k_{1},k_{2}) = P(k_{1},k_{2}) / P_{r}(k_{1},k_{2})$ of the cellular metabolic network of E. \emph{coli}~\cite{jeong1} (known to be a fractal network) and for the Internet at the router level~\cite{burch} (a non-fractal network), where $P_{r}(k_{1},k_{2})$ is obtained by randomly swapping the links without modifying the degree distribution.

Fig.~\ref{kk} shows a dramatic difference between the two networks. The network of E. \emph{coli}, that is a fractal network, present an anti-correlation of the degrees (or dissasortativity~\cite{newman1,newman2}), which means that mostly high degree nodes are linked to low degree nodes. This property leads to fractal networks. On the other hand, the Internet exhibits a high correlation between degrees leading to a non-fractal network.

With this idea in mind, in 2006 C. Song, S. Havlin and H. Makse presented a model that elucidates the way new nodes must be connected to the old
ones in order to build a fractal, a non-fractal network, or a mixture between fractal and non-fractal network~\cite{song2}. This model shows that, indeed, the correlations between degrees of the nodes are a determinant factor for the fractality of a network.
This model was later extended ~\cite{lazaros2} to allow loops in the network, while preserving the self-similarity and fractality properties.

\begin{figure}
\centering{ { \resizebox{6cm}{!} {\includegraphics{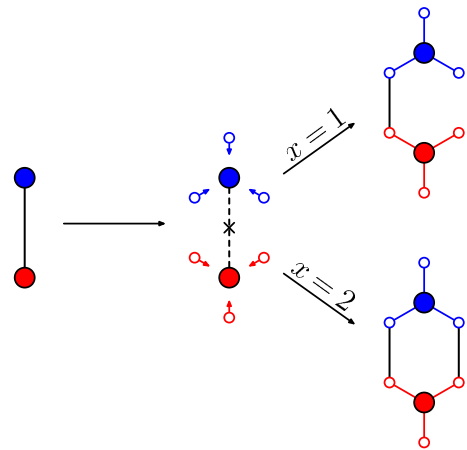}}}
}
\caption{\label{FIG_draw_model} The model grows from a small network, usually two nodes connected to each other.
During each step and for every link in the system, each endpoint of a link produces $m$ offspring nodes (in this drawing $m=3$).
In this case, with probability $e=1$ the original link is removed and $x$
new links between randomly selected nodes of the new generation are added.
Notice that the case of $x=1$ results in a tree structure, while loops
appear for $x>1$.
}
\end{figure}

The algorithm is as follows (see Fig.~\ref{FIG_draw_model}): In generation $n=0$, start with two nodes connected by one link. Then, generation $n+1$ is obtained recursively by attaching $m$ new nodes to the endpoints of each link $l$ of generation $n$. In addition, with probability $e$ remove link $l$ and add $x$ new links connecting pairs of new nodes attached to the endpoints of $l$. 

The degree distribution, diameter and fractal dimension can be easily calculated. For example, if $e=1$ (pure fractal network), the degree distribution follows a power-law $P(k) \sim k^{-\gamma}$ with exponent $\gamma = 1+\textrm{log}(2m+x)/\textrm{log}~m$ and the fractal dimension is $d_B=\textrm{log}(2m+x)/\textrm{log}~m$. The diameter $L$ scales, in this case, as power of the number of nodes as $L \sim N^{1/d_{B}}$~\cite{song2,song3}. Later, in Section ``Properties of Fractal and Transfractal Networks'', several topological properties are shown for this model network.

\subsection{$(u,v)$-flowers}

\begin{figure}[ht]
  \vspace*{0.3cm}\includegraphics*[width=0.40\textwidth]{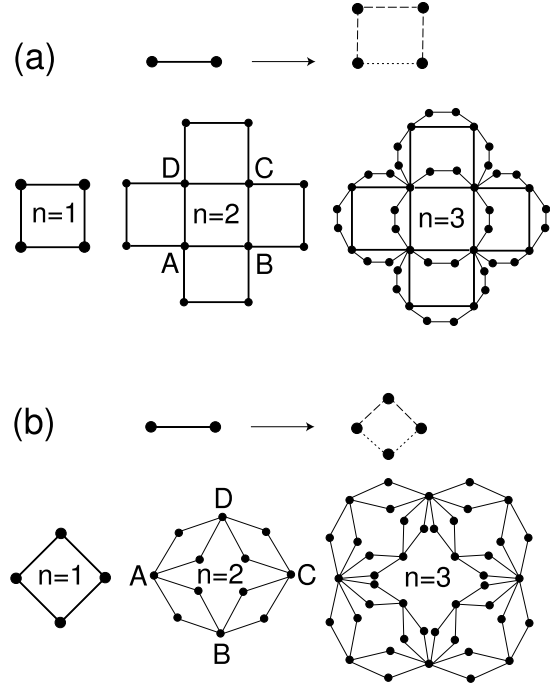}
\caption{$(u,v)$-flowers with $u+v=4$ ($\gamma=3$).
(a)~$u=1$ (dotted line) and $v=3$ (broken line). (b)~$u=2$ and $v=2$.  The graphs may also
be iterated by joining four replicas of generation $n$ at the hubs A and B, for (a), or A and C, for~(b).
}
\label{g3}
\end{figure}

In 2006, H. Rozenfeld, S. Havlin and D. ben-Avraham proposed a new family of recursive \emph{deterministic} scale-free networks, the $(u,v)$-\emph{flowers}, that generalize both, the original scale-free model of Berker and Ostlund~\cite{berker1} and the pseudo-fractal network of Dorogovstev, Goltsev and Mendes~\cite{dorogovtsev} and that, by appropriately varying its two parameters $u$ and $v$, leads to either fractal networks or non-fractal networks~\cite{rozenfeld1, rozenfeld2}. The algorithm to build the $(u,v)$-flowers is the following: In generation $n=1$ one starts with a cycle graph (a ring) consisting of $u+v\equiv w$ links and nodes (other choices are possible). Then, generation $n+1$ is obtained recursively by replacing each link by two parallel paths of $u$ and $v$ links long.  Without loss of generality, $u \leq v$.  Examples of $(1,3)$-
and $(2,2)$-flowers are shown in Fig.~\ref{g3}.  The DGM network corresponds to the special case of $u=1$ and $v=2$ and the Berker and Ostlund model corresponds to $u=2$ and $v=2$.

An essential property of the $(u,v)$-flowers is that they are self-similar, as evident from an equivalent method of construction: to produce generation $n+1$, make $w=u+v$ copies of the net in generation $n$ and join them at the hubs.

The number of links of a $(u,v)$-flower of generation $n$ is
\begin{equation}
\label{Mn}
M_n=(u+v)^n= w^n,
\end{equation}
and the number of nodes is
\begin{equation}
\label{Nn}
N_n=\Big(\frac{w-2}{w-1}\Big)w^n+\Big(\frac{w}{w-1}\Big)\,.
\end{equation}

The degree distribution of the $(u,v)$-flowers can also be easily obtained since by construction, $(u,v)$-flowers have only nodes of degree $k=2^m$, $m=1,2,\dots,n$. As in the DGM case, $(u,v)$-flowers follow a scale-free degree distribution, $P(k)\sim k^{-\gamma}$,
of degree exponent
\begin{equation}
\label{gamma}
\gamma=1+\frac{\ln(u+v)}{\ln2}\,.
\end{equation}

\begin{figure}[ht]
  \vspace*{0.3cm}\includegraphics*[width=0.40\textwidth]{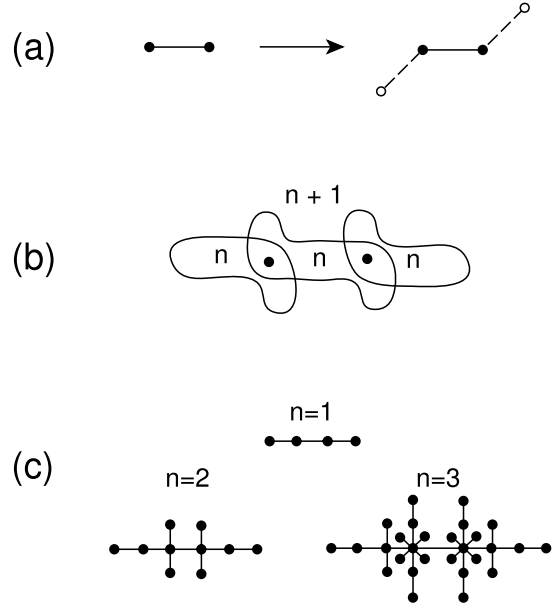}
\caption{The $(1,2)$-tree. 
(a)~Each link in generation $n$ is replaced by a chain of $u=1$ links, to which ends one attaches chains of $v/2=1$ links. (b)~Alternative method of construction highlighting self-similarity: $u+v=3$ replicas of generation $n$ are joined at the hubs.
(c)~Generations $n=1,2,3$.
}
\label{tree}
\end{figure}

Recursive scale-free {\it trees\/} may be defined in analogy to the flower nets.  If $v$ is even, one obtains generation 
$n+1$ of a $(u,v)$-tree by replacing every link in generation $n$ with a chain of $u$ links, and attaching
to each of its endpoints chains of $v/2$ links. Fig.~\ref{tree} shows how this works for the $(1,2)$-tree.
If $v$ is odd, attach to the endpoints (of the chain of $u$ links) chains of length $(v\pm1)/2$.  The trees may be also constructed by successively joining $w$ replicas at the appropriate hubs, and they too are self-similar.  They share many of the fundamental scaling properties with 
$(u,v)$-flowers: Their degree distribution is also scale-free, with the same degree exponent as $(u,v)$-flowers.

The self-similarity of $(u,v)$-flowers, coupled with the fact that different replicas meet at a {\it single\/} 
node, makes them amenable to exact analysis by renormalization techniques.  The lack of loops, in the case of $(u,v)$-trees, further simplifies their analysis~\cite{berker1,rozenfeld1,rozenfeld2, bollt}.

\subsubsection{Dimensionality of the $(u,v)$-flowers}
There is a vast difference between $(u,v)$-nets with $u=1$ and $u>1$.
If $u=1$ the diameter $L_n$ of the $n$-th generation flower scales linearly with $n$.  For example, $L_n$ for the $(1,2)$-flower~\cite{dorogovtsev} and
$L_n=2n$ for the $(1,3)$-flower.  It is easy to see that the diameter of the $(1 ,v)$-flower, for $v$ odd, is
$L_n=(v-1)n+(3-v)/2$, and, in general one can show that $L_n\sim(v-1)n$.

For $u>1$, however, the diameter grows as a power of $n$.  For example, for the $(2,2)$-flower we find
$L_n=2^n$, and, more generally, the diameter satisfies $L_n\sim u^n$.
To summarize,
\begin{equation}
\label{Ln}
L_n \sim \left\{ \begin{array}{ll}
(v-1)n &  u=1,\\
u^{n} & u > 1,
\end{array} \right.\qquad{\rm flowers.}
\end{equation}
Similar results are quite obvious for the case of $(u,v)$-trees, where \begin{equation}
\label{Ln_trees}
L_n \sim \left\{ \begin{array}{ll}
vn &  u=1,\\
u^{n} & u > 1,
\end{array} \right.\qquad{\rm trees.}
\end{equation}

Since $N_n\sim (u+v)^n$ [Eq.~(\ref{Nn})], we can recast these relations as
\begin{equation}
\label{L}
L \sim \left\{ \begin{array}{ll}
\ln N &  u=1,\\
N^{\ln u/\ln(u+v)} & u > 1.
\end{array} \right.
\end{equation}
Thus, $(u,v)$-nets are {\it small world\/} only in the case of $u=1$.  For $u>1$, the diameter increases as a power of $N$, just as in {\it finite\/}-dimensional objects, and the nets are in fact {\it fractal\/}. 
For $u>1$, the change of mass upon the rescaling of length by a factor $b$ is
\begin{equation}
\label{defdf}
N(bL)=b^{d_B}N(L)\,,
\end{equation}
where $d_B$ is the fractal dimension~\cite{bahbook}.  In this case,  $N(uL)=(u+v)N(L)$, so
\begin{equation}
\label{df1}
d_B=\frac{\ln(u+v)}{\ln u}\,,\qquad u>1\,.
\end{equation}
 
\subsubsection{Transfinite Fractals}
Small world nets, such as $(1,v)$-nets, are {\it infinite\/}-dimensional.  Indeed, their mass ($N$, or $M$) increases faster than any power (dimension) of their diameter.  Also, note that a naive application of (\ref{df}) to $u\to1$ yields $\df\to\infty$. In the case of $(1,v)$-nets one can use their weak self-similarity to define a new measure of dimensionality, $\tdf$, characterizing how mass scales with diameter:
\begin{equation}
\label{deftdf}
N(L+\ell)=e^{\ell\tdf}N(L)\,.
\end{equation}
Instead of a multiplicative rescaling of length, $L\mapsto bL$, a slower additive mapping, $L\mapsto L+\ell$, that reflects the small world property is considered. Because the exponent $\tdf$ usefully distinguishes between different graphs of infinite dimensionality, $\tdf$ has been termed the {\it transfinite\/} fractal dimension of the network.  Accordingly, objects that are self-similar and have infinite dimension (but finite transfinite dimension), such as the $(1,v)$-nets, are termed transfinite fractals, or {\it transfractals\/}, for short.

For $(1,v)$-nets, we see that upon `zooming in' one generation level the mass increases by a factor
of $w=1+v$, while the diameter grows from $L$ to  $L+v-1$ (for flowers), or to $L+v$ (trees).  Hence their transfractal dimension is
\begin{equation}
\label{tdf}
\tdf= \left\{ \begin{array}{ll}
\frac{\ln(1+v)}{v} &(1,v)\text{-trees,}\\ \\
\frac{\ln(1+v)}{v-1} &(1,v)\text{-flowers.}
\end{array} \right.
\end{equation}

There is some arbitrariness in the selection of $e$ as the base of the exponential in the definition (\ref{deftdf}). However the base is inconsequential for the sake of comparison between dimensionalities of different objects.  Also, {\it scaling relations\/} between various transfinite exponents hold, irrespective of the choice of base: consider the scaling relation of Eq.~\ref{relation}
valid for fractal scale-free nets of degree exponent $\gamma$~\cite{song,song2}.  
For example, in the fractal $(u,v)$-nets (with $u>1$) renormalization reduces lengths by a factor $b=u$ and all degrees are reduced by a factor of 2, so $b^{\dk}=2$. Thus $\dk=\ln2/\ln u$, and since $d_B=\ln(u+v)/\ln u$ and $\gamma=1+\ln(u+v)/\ln2$, as discussed above, 
the relation (\ref{relation}) is indeed satisfied.

For transfractals,  renormalization reduces distances by an {\it additive\/}
length, $\ell$, and we express the self-similarity manifest in the degree distribution as
\begin{equation}
\label{scaletPk}
P'(k)=e^{\ell\tdk}P(e^{-\ell\tdk}k)\,,
\end{equation}
where $\tdk$ is the transfinite exponent analogous to $\dk$.  Renormalization of the transfractal $(1,v)$-nets
reduces the link lengths by $\ell=v-1$ (for flowers), or $\ell=v$ (trees), while all degrees are halved.
Thus,
\[
 \tdk= \left\{ \begin{array}{ll}
\frac{\ln2}{v} &(1,v)\text{-trees,}\\ \\
\frac{\ln2}{v-1} &(1,v)\text{-flowers.}
\end{array} \right.
\]
Along with (\ref{tdf}), this result confirms that the scaling relation  
\begin{equation}
\label{gtfrac}
\gamma=1+\frac{\tdf}{\tdk}
\end{equation}
is valid also for transfractals, and regardless of the choice of base.
A general proof of this relation is practically identical to the proof of~(\ref{gfrac})~\cite{song}, merely replacing fractal with transfractal scaling throughout the argument.  

For scale-free transfractals, following $m=L/\ell$ renormalizations
the diameter and mass reduce to order one, and the scaling (\ref{deftdf}) implies $L\sim m\ell$, $N\sim e^{m\ell\tdf}$, so that
\[
L\sim \frac{1}{\tdf}\ln N\,,
\]
in accordance with their small world property.  At the same time the scaling (\ref{scaletPk}) implies
$K\sim e^{m\ell\tdk}$, or $K\sim N^{\tdk/\tdf}$.  Using the scaling relation (\ref{gtfrac}), we rederive 
$K\sim N^{1/(\gamma-1)}$, which is indeed valid for scale-free nets {\it in general\/}, be they fractal or transfractal.

\section{Properties of Fractal and Transfractal Networks}

The existence of fractality in complex networks immediately calls for the question
of what is the importance of such a structure in terms of network properties.
In general, most of the relevant applications seem to be modified to a larger or
lesser extent, so that fractal networks can be considered to form a separate network sub-class,
sharing the main properties resulting from the wide distribution of regular scale-free networks,
but at the same time bearing novel properties.
Moreover, from a practical point of view a fractal network
can be usually more amenable to analytic treatment.

In this section we summarize some of the applications that seem to distinguish
fractal from non-fractal networks.

\subsection{Modularity}

Modularity is a property closely related to fractality. Although this term does not
have a unique well-defined definition we can claim that modularity refers to
the existence of areas in the network where groups of nodes share some common
characteristics, such as preferentially connecting within this area (the `module')
rather than to the rest of the network. The
isolation of modules into distinct areas is a complicated task and in most cases
there are many possible ways (and algorithms) to partition a network into modules.

Although networks with significant degree of modularity are not necessarily
fractals, practically all fractal networks are highly modular in structure.
Modularity naturally emerges from the effective `repulsion' between hubs.
Since the hubs are not directly connected to each other, they usually dominate
their neighborhood and can be considered as the `center of mass' for a given module.
The nodes surrounding hubs are usually assigned to this module.

The renormalization property of self-similar networks is very useful for estimating
how modular a given network is, and especially for how this property is modified
under varying scales of observation. We can use a simple definition for
modularity $M$, based on the idea that 
the number of links connecting nodes within a module, $L_i^{\rm in}$, is higher than the number of link connecting nodes in different modules, $L_i^{\rm out}$.
For this purpose, the boxes that result from the box-covering
method at a given length-scale $\ell_B$ are identified as the network modules for
this scale. This partitioning assumes that the minimization of the number of boxes
corresponds to an increase of modularity, taking advantage of the idea that all
nodes within a box can reach each other within less than $\ell_B$ steps. 
This
constraint tends to assign the largest possible number of nodes in a given
neighborhood within the same box, resulting in an optimized modularity function.

A definition of the modularity function $M$ that takes advantage of the special features
of the renormalization process is, thus, the following~\cite{lazaros2}:
\begin{equation}
\label{EQ_modularity}
M(\ell_B) = \frac{1}{N_B} \sum_{i=1}^{N_B} \frac{L_i^{\rm in}}{L_i^{\rm out}} \,,
\end{equation}
where the sum is over all the boxes.

The value of $M$ through Eq. (\ref{EQ_modularity}) for a given $\ell_B$ value
is of small usefuleness on its own, though. We can gather more information on the
network structure if we measure $M$ for different values of $\ell_B$. If the dependence
of $M$ on $\ell_B$ has the form of a power-law, as if often the case in practice, then
we can define the modularity exponent $d_M$ through
\begin{equation}
\label{EQ_modul}
M(\ell_B) \sim \ell_B^{d_M} \,.
\end{equation}
The exponent $d_M$ carries the important information of how modularity scales with the length,
and separates modular from non-modular networks. The value of $d_M$ is easy to compute
in a $d$-dimensional lattice, since the number of links within any module scales with its
bulk, as $L_i^{\rm in}\sim \ell_B^d$  and the number of links outside the module scale
with the length of its interface, i.e. $L_i^{\rm out}\sim \ell_B^{d-1}$. So, the resulting
scaling is $M\sim \ell_B$ i.e. $d_M=1$. This is also the borderline value that separates
non-modular structures ($d_M<1$) from modular ones ($d_M>1$).

For the Song-Havlin-Makse fractal model introduced in the previous section, a module can be identified
as the neighborhood around a central hub. In the simplest version with $x=1$,
the network is a tree, with well-defined modules. Larger values of $x$ mean that
a larger number of links are connecting different modules, creating more loops and
`blurring' the discreteness of the modules,  so that we can vary the degree of modularity in the network.
For this model, it is also possible to analytically calculate the value of the exponent $d_M$.

During the growth process at step $t$, the diameter in the network model increases multiplicatively
as $L(t+1)=3L(t)$. The number of links within a module grows with
$2m+x$ (each node on the side of one link gives rise to $m$ new links
and $x$ extra links connect the new nodes), while the number of links pointing out of a module is by definition
proportional to $x$. Thus, the modularity
$M(\ell_B)$ of a network is proportional to $(2m+x)/x$. Eq. (\ref{EQ_modul}) can
then be used to calculate $d_M$ for the model:
\begin{equation}
\frac{2m+x}{x} \sim 3^{d_M} \,,
\end{equation}
which finally yields
\begin{equation}
d_M = \frac{\ln\left(2\frac{m}{x}+1\right)}{\ln 3} \,.
\end{equation}
So, in this model the important quantity that determines the degree of modularity in the system
is the ratio of the growth parameters $m/x$.

Most of the real-life networks that have been measured display some sort of modular character,
i.e. $d_M>1$, although many of them have values very close to 1. Only in a few cases
we have observed exponents $d_M<1$. Most interesting is, though, the case of
$d_M$ values much larger than 1, where a large degree of modularity is observed and this
trend is more pronounced for larger length-scales.

The importance of modularity as described above can be demonstrated in biological networks.
There, it has been suggested that the boxes may correspond to functional modules
and in protein interaction networks, for example, there may be an evolution drive of the system
behind the development of its modular structure.

\subsection{Robustness}

Shortly after the discovery of the scale-free property, the first important
application of their structure was perhaps their extreme resilience to removal of
random nodes~\cite{cohen1,cohen3,albert1,beygelzimer,rozenfeld2,serrano1,serrano2}. At the same time such a network was found to be quite vulnerable
to an intentional attack, where nodes are removed according to descreasing
order of their degree~\cite{cohen2,lazaros3}. The resilience of a network is usually quantified through
the size of the largest remaining connected cluster $S_{max}(p)$, when a fraction $p$ of
the nodes has been removed according to a given strategy. At a critical point $p_c$
where this size becomes equal to $S_{max}(p_c)\simeq 0$, we consider that the network
has been completely disintegrated. For the random removal case, this threshold is
$p_c\simeq 1$, i.e. practically all nodes need to be destroyed. In striking contrast,
for intentional attacks $p_c$ is in general of the order of only a few percent, although
the exact value depends on the system details.

\begin{figure}
\centering{ { \resizebox{8.8cm}{!} {\includegraphics{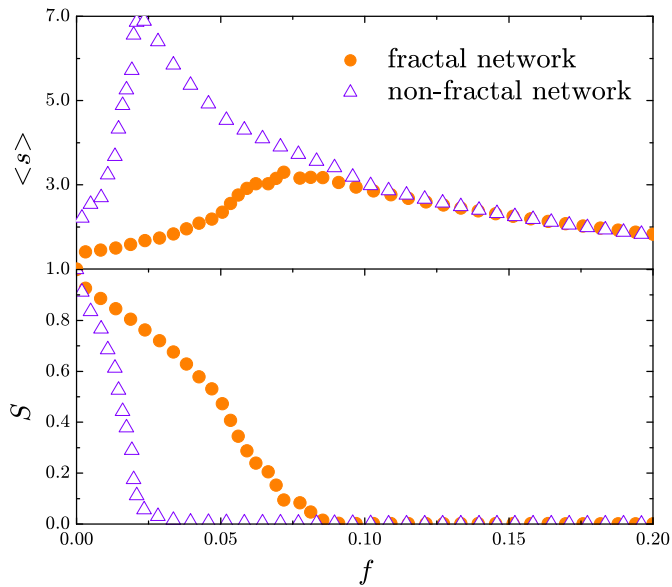}}}
}
\caption{\label{FIG_Smax} Vulnerability under intentional attack of a
non-fractal Song-Makse-Havlin network (for $e=0$) and a fractal Song-Makse-Havlin network (for $e=1$). The  plot shows the relative size of the largest cluster, $S$, and the average size of the remaining
isolated clusters, $\langle s \rangle$ as a function of the removal fraction $f$ of the largest hubs for both networks.
}
\end{figure}

Fractality in networks considerably strengthens the robustness against intentional attacks,
compared to non-fractal networks with the same degree exponent $\gamma$.
In Fig. \ref{FIG_Smax} the comparison between two such networks clearly shows that the critical fraction
$p_c$ increases almost 4 times from $p_c\simeq 0.02$ (non-fractal topology) to $p_c\simeq 0.09$
(fractal topology). These networks have the same $\gamma$ exponent, the same number of links,
number of nodes, number of loops and the same clustering coefficient, differing only in whether
hubs are directly connected to each other. The fractal property, thus, provides a way of increasing
resistance against the network collapse, in the case of a targeted attack.

The main reason behind this behavior is the dispersion of hubs in the network. A hub is usually
a central node that helps other nodes to connect to the main body of the system. When the hubs are
directly connected to each other, this central core is easy to destroy in a targeted attack
leading to a rapid collapse of the network. On the contrary, isolating the hubs into different
areas helps the network to retain connectivity for longer time, since destroying the hubs now
is not similarly catastrophic, with most of the nodes finding alternative paths through other connections.

The advantage of increased robustness derived from the combination of modular and fractal
network character, may provide valuable hints on why most biological networks have evolved towards
a fractal architecture (better chance of survival against lethal attacks).

\subsection{Degree Correlations}

We have already mentioned the importance of hub-hub correlations or anti-correlations in fractality.
Generalizing this idea to nodes of any degree, we can ask what is the joint degree probability
$P(k_1,k_2)$ that a randomly chosen link connects two nodes with degree $k_1$ and $k_2$, respectively.
Obviously, this is a meaningful question only for networks with a wide degree distribution, otherwise
the answer is more or less trivial with all nodes having similar degrees. A similar and perhaps more
useful quantity is the conditional degree probability $P(k_1|k_2)$, defined as the probability that
a random link from a node having degree $k_2$ points to a node with degree $k_1$. In general, the
following balance condition is satisfied
\begin{equation}
k_2 P(k_1|k_2) P(k_2) = k_1 P(k_2|k_1) P(k_1) \,.
\end{equation}
It is quite straightforward to calculate $P(k_1|k_2)$ for completely uncorrelated networks.
In this case, $P(k_1|k_2)$ does not depend on $k_2$, and the probability to chose a node with degree $k_1$
becomes simply $P(k_1|k_2)=k_1 P(k_1) / \langle k_1 \rangle$. In the case where degree-degree correlations
are present, though, the calculation of this function is very difficult, even when restricting
ourselves to a direct numerical evaluation, due to the emergence of huge fluctuations.

We can still estimate this function, though, using again the self-similarity principle.
If we consider that the function $P(k_1,k_2)$ remains invariant under the network renormalization scheme
described above, then it is possible to show that
\begin{equation}
P(k_1,k_2) \sim k_1^{-(\gamma-1)}k_2^{-\epsilon} \,\,\, (k_1>k_2) \,,
\end{equation}
and similarly
\begin{equation}
P(k_1|k_2) \sim k_1^{-(\gamma-1)}k_2^{-(\epsilon-\gamma+1)} \,\,\, (k_1>k_2) \,,
\end{equation}
In the above equations we have also introduced the correlation exponent $\epsilon$,
which characterizes the degree of correlations in a network. For example, the case of
uncorrelated networks is described by the value $\epsilon=\gamma-1$.

The exponent $\epsilon$ can be measured quite accurately using an appropriate quantity.
For this purpose, we can introduce a measure such as
\begin{equation}
E_b(k)\equiv \frac{\int_{bk}^{\infty} P(k|k_2)dk_2}
{\int_{bk}^{\infty} P(k)dk} \,,
\label{EQdefEbk}
\end{equation}
which estimates the probability that a node with degree $k$ has neighbors with
degree larger than $bk$, and $b$ is an arbitrary parameter that has been shown not
to influence the results. It is easy to show that
\begin{equation}
E_b(k) \sim \frac{k^{1-\epsilon}}{k^{1-\gamma}} = k^{-(\epsilon-\gamma)} \,.
\end{equation}
This relation allows us to estimate $\epsilon$ for a given network, after calculating
the quantity $E_b(k)$ as a function of $k$.

The above discussion can be equally applied to both fractal and non-fractal networks.
If we restrict ourselves to fractal networks, then we can develop our theory a bit further.
If we consider the probability ${\cal E}(\ell_B)$ that the
largest degree node in each box is connected directly with the other largest degree nodes
in other boxes (after optimally covering the network), then this quantity scales as a power
law with $\ell_B$:
\begin{equation}
{\cal E} (\ell_B) \sim \ell_B^{-d_e} \,,
\end{equation}
where $d_e$ is a new exponent describing the probability of hub-hub connection~\cite{song2}.
The exponent $\epsilon$, which describes correlations
over any degree, is related to $d_e$, which refers to correlations between hubs only. The resulting
relation is
\begin{equation}
\epsilon = 2+d_e/d_k = 2+(\gamma-1)\frac{d_e}{d_B} \,.
\end{equation}
For an infinite fractal
dimension $d_B\to \infty$, which is the onset of non-fractal networks that cannot be
described by the above arguments, we have the limiting case of $\epsilon=2$. This value
separates fractal from non-fractal networks, so that fractality is indicated by $\epsilon>2$.
Also, we have seen that the line $\epsilon=\gamma-1$ describes networks for which correlations
are minimal. Measurements of many real-life networks have verified the above statements,
where networks with $\epsilon>2$ having been clearly characterized as fractals with alternate methods.
All non-fractal networks have values of $\epsilon<2$ and the distance from the
$\epsilon=\gamma-1$ line determines how much stronger or weaker the correlations are, compared to the uncorrelated case.

In short, using the self-similarity principle makes it possible to gain a lot of insight
on network correlations, a notoriously difficult task otherwise. Furthermore, the
study of correlations can be reduced to the calculation of a single exponent $\epsilon$, 
which is though capable of delivering a wealth of information on the network topological
properties.

\subsection{Diffusion and Resistance}

Scale-free networks have been described as objects of infinite dimensionality.
For a regular structure this statement would suggest that one can simply use
the known diffusion laws for $d=\infty$. Diffusion on scale-free structures, however, is
much harder to study, mainly due to the lack of translational symmetry in the system
and different local environments. Although exact results are still not available,
the scaling theory on fractal networks provides the tools to better understand processes,
such as diffusion and electric resistance.

In the following, we describe diffusion through the average first-passage time $T_{AB}$, which is the
average time for a diffusing particle to travel from node A to node B. At the same time, assuming that each link in the network has an electrical resistance of 1 unit, we can describe the electrical properties through the resistance between the two nodes A and B, $R_{AB}$. 

The connection between diffusion (first-passage time) and electric networks has long been established in homogeneous systems. This connection is usually expressed through the Einstein relation~\cite{bahbook}. The Einstein relation is of great importance because it connects a \emph{static quantity} $R_{AB}$ with a \emph{dynamic quantity} $T_{AB}$. In other words, the behavior of a diffusing particle can be inferred by simply having knowledge of a static topological property of the network.

In any renormalizable network the scaling of $T$ and $R$ follow the form:
\begin{equation}
\label{EQ_expon}
T'/T = \ell_B^{-d_w}, ~ R'/R = \ell_B^{-\zeta},
\end{equation}
where $T'$ ($R'$) and $T$ ($R$) are the first-passage time (resistance) for the renormalized and original networks, respectively. The dynamical exponents $d_w$ and $\zeta$ characterize the scaling in any lattice or network that remains invariant under renormalization. The Einstein relation relates these two exponents through the dimensionality of the substrate $d_B$, according to:
\begin{equation}
d_w = \zeta + d_B \,.
\end{equation}

The validity of this relation in inhomogeneous complex networks, however, is not yet clear. Still, in fractal
and transfractal networks there are many cases where this relation has been proved to be valid,
hinting towards a wider applicability. For example, in Refs.~\cite{bollt,rozenfeld1} it has been shown that the Einstein Relation~\cite{bahbook} in $(u,v)$-flowers and $(u,v)$-trees is valid for any $u$ and $v$, that is for both fractal and transfractal networks. In general, in terms of the scaling theory we can study diffusion and resistance (or conductance) in a similar manner~\cite{lazaros2}.

Because of the highly inhomogeneous character of the structure, though, we are interested in how these
quantities behave as a function of the end-node degrees $k_1$ and $k_2$ when they are separated by a given distance $\ell$. Thus, we are looking for the full dependence of $T(\ell;k_1,k_2)$ and $R(\ell;k_1,k_2)$. Obviously, for lattices or networks with narrow degree distribution there is no degree dependence and those results should be a function of $\ell$ only.

\begin{figure}
\centering{
(a) {\resizebox{4.2cm}{!} {\includegraphics{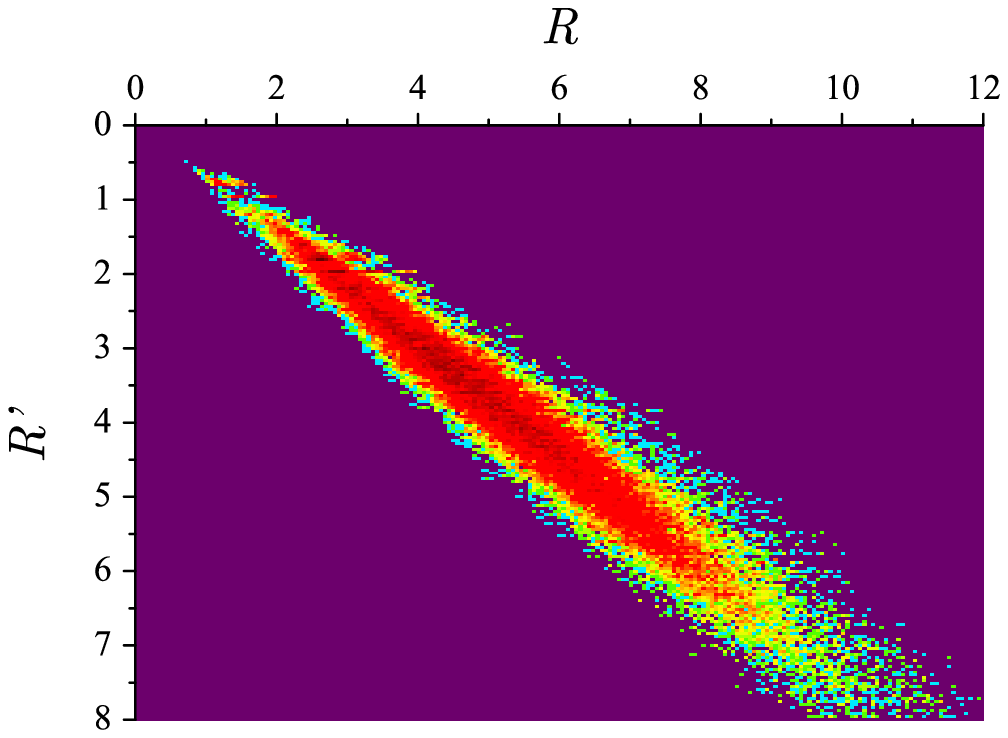}}}
(b){\resizebox{4.2cm}{!} {\includegraphics{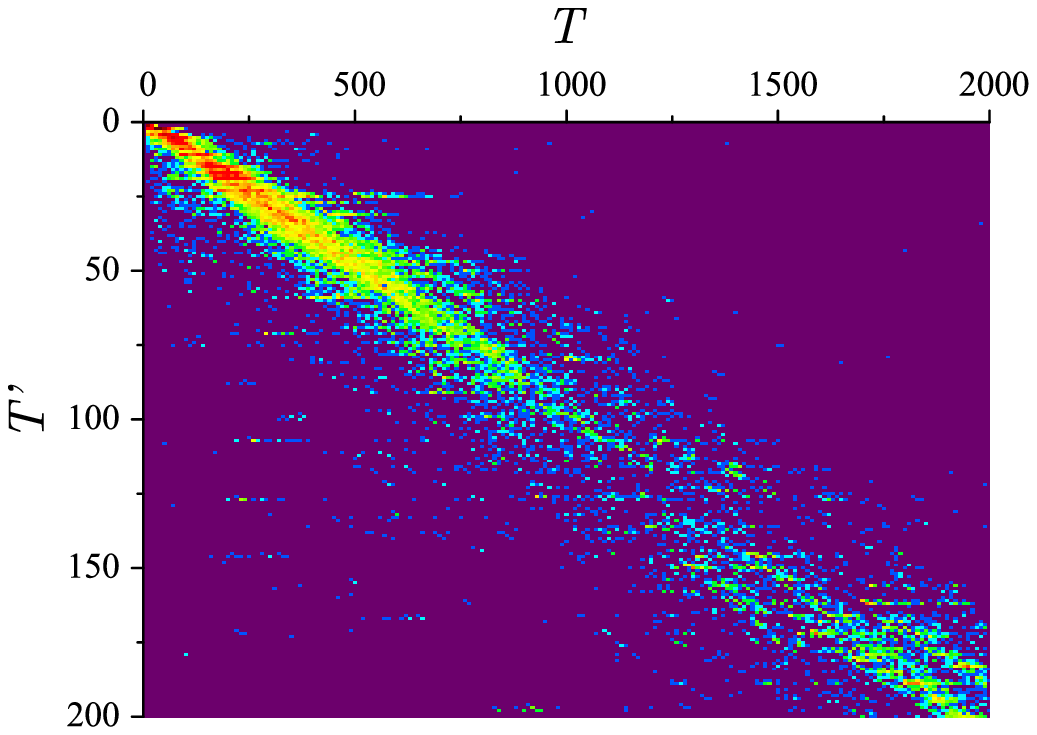}}}
}
\caption{\label{FIGrenor1} Typical behavior of the probability distributions for the resistance $R$ vs $R'$ and the diffusion time $T$ vs $T'$, respectively,
for a given $\ell_B$ value. Similar plots for other $\ell_B$ values verify
that the ratios of these quantities during a renormalization stage are roughly constant for all pairs of nodes in a given biological network.
}
\end{figure}

\begin{figure}
\centering{ { \resizebox{8.8cm}{!} {\includegraphics{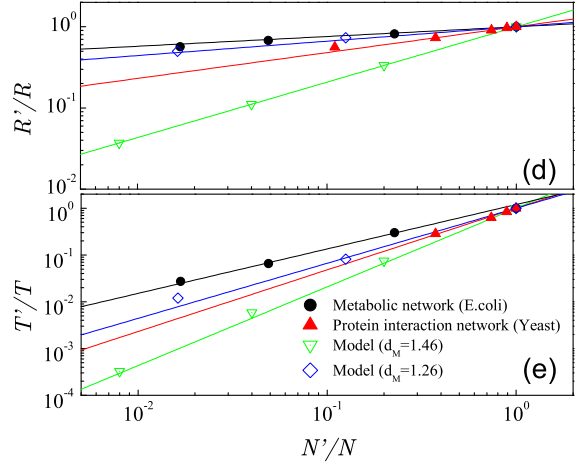}}}
}
\caption{\label{FIGrenor2}
Average value of the ratio of resistances $R/R'$ and diffusion times $T/T'$, as measured for different $\ell_B$ values (each point corresponds to a different value of $\ell_B$).
Results are presented for both biological networks, and two fractal network models with different $d_M$ values.
The slopes of the curves correspond to the exponents $\zeta/d_B$ (top panel) and $d_w/d_B$ (bottom panel).
}
\end{figure}

For self-similar networks, we can rewrite Eq.~(\ref{EQ_expon}) above as
\begin{equation}
\label{EQ_ratios}
\frac{T'}{T} = \left( \frac{N'}{N} \right)^{d_w/d_B} \,,\, \frac{R'}{R} = \left( \frac{N'}{N} \right)^{\zeta/d_B} \,.
\end{equation}
where we have taken into account Eq.~(\ref{dh}). This approach offers the practical advantage that the variation
of $N'/N$ is larger than the variation of $\ell_B$, so that the exponents calculation can be more accurate.
To calculate these exponenets, we fix the box size $\ell_B$ and we measure the diffusion time $T$ and resistance $R$ between any
two points in a network before and after renormalization. If for every such pair we plot the
corresponding times and resistances in $T'$ vs $T$ and $R'$ vs $R$ plots, as shown in Fig.~\ref{FIGrenor1}, then all these points fall in a narrow area, suggesting a constant value for the ratio $T'/T$ over the entire network. Repeating this procedure for different $\ell_B$ values yields other ratio values.
The plot of these ratios vs $N'/N$ (Fig.~\ref{FIGrenor2}) finally exhibits a power-law dependence, verifying
Eq.~(\ref{EQ_ratios}). We can then easily calculate the exponents $d_w$ and $\zeta$ from the slopes
in the plot, since the $d_B$
exponent is already known through the standard box-covering methods. It has been shown that the results
for many different networks are consistent, within statistical error, with the Einstein relation~\cite{rozenfeld1,lazaros2}.

\begin{figure}
\centering{ { \resizebox{8.8cm}{!} {\includegraphics{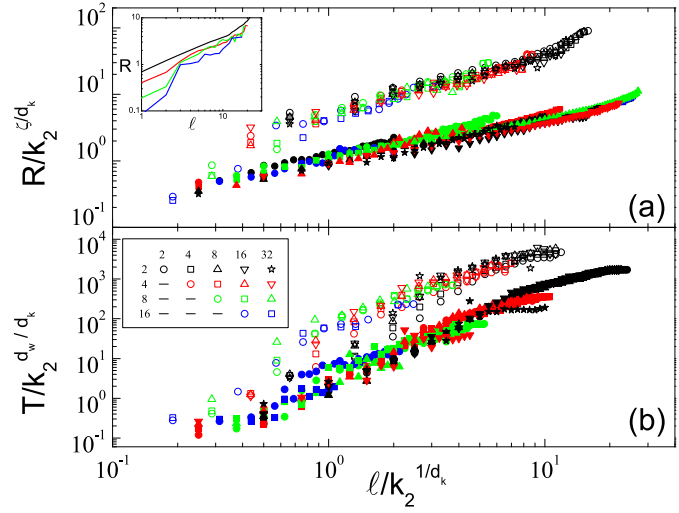}}}
}
\caption{\label{FIG_rw_res_rescaled} Rescaling of (a) the resistance and (b) the diffusion time
according to Eqs.~(\protect\ref{EQfinalscaling}) and (\protect\ref{EQfinalTscaling})
for the protein interaction network of yeast
(upper symbols) and the Song-Havlin-Makse model for $e=1$ (lower filled symbols).
The data for PIN have been vertically shifted upwards by one decade for clarity.
Each symbol corresponds to a fixed ratio $k_1/k_2$ and the different colors denote
a different value for $k_1$. Inset: Resistance $R$ as a function of distance $\ell$,
before rescaling, for constant ratio $k_1/k_2=1$ and different $k_1$ values.
}
\end{figure}

The dependence on the degrees $k_1$, $k_2$ and the distance $\ell$ can also be calculated in a scaling form
using the self-similarity properties of fractal networks. After renormalization, a node with degree $k$
in a given network, will have a degree $k' = \ell_B^{-d_k} k$ according to Eq.~(\ref{sl}). At the same time
all distances $\ell$ are scaled down according to $\ell'=\ell/\ell_B$. This means that Eqs.(\ref{EQ_ratios})
can be written as
\begin{eqnarray}
\label{EQrrr}
R'(\ell' ; k_1',k_2') = \ell_B^{-\zeta} R(\ell ; k_1,k_2) \\
T'(\ell' ; k_1',k_2') = \ell_B^{-d_w} T(\ell ; k_1,k_2) \,.
\end{eqnarray}
Substituting the renormalized quantities we get:
\begin{equation}
R' ( \ell_B^{-1}\ell ; \ell_B^{-d_k} k_1, \ell_B^{-d_k}k_2) = \ell_B^{-\zeta} R(\ell ; k_1,k_2) \,.
\end{equation}
The above equation holds for all values of $\ell_B$, so we can select this quantity to be
$\ell_B=k_2^{1/d_k}$. This constraint allows us to reduce the number of variables in the equation,
with the final result:
\begin{equation}
\label{EQrescaleR}
R\left( \frac{\ell}{k_2^{1/d_k}} ; \frac{k_1}{k_2}, 1 \right) = k_2^{-\zeta/d_k} R( \ell_B ; k_1,k_2 ) \,.
\end{equation}
This equation suggests a scaling for the resistance $R$:
\begin{equation}
\label{EQfinalscaling}
R(\ell ; k_1,k_2) = k_2^{\zeta/d_k} f_R\left( \frac{\ell}{k_2^{1/d_k}} , \frac{k_1}{k_2}\right) \,,
\end{equation}
where $f_R()$ is an undetermined function. All the above arguments
can be repeated for the diffusion time, with a similar expression:
\begin{equation}
\label{EQfinalTscaling}
T(\ell ; k_1,k_2) = k_2^{d_w/d_k} f_T\left( \frac{\ell}{k_2^{1/d_k}} , \frac{k_1}{k_2}\right) \,,
\end{equation}
where the form of the right-hand function may be different.
The final result for the scaling form is Eqs.~(\ref{EQfinalscaling}) and (\ref{EQfinalTscaling}),
which is also supported by the numerical data collapse in Fig.~\ref{FIG_rw_res_rescaled}.
Notice that in the case of homogeneous networks, where there is almost no $k$-dependence, the unknown functions in the rhs
reduce to the forms $f_R(x,1)=x^\zeta$, $f_T(x,1)=x^{d_w}$, leading to the well-established classical relations $R\sim\ell^\zeta$ and $T\sim\ell^{d_w}$.

\section*{Future Directions}

Fractal networks combine features met in fractal geometry and in network theory. As such, they
present many unique aspects. Many of their properties have been well-studied and understood,
but there is still a great amount of open and unexplored questions remaining to be studied.

Concerning the structural aspects of fractal networks, we have described that 
in most networks the degree distribution $P(k)$, the joint degree distribution
$P(k_1,k_2)$ and a number of other quantities remain invariant under renormalization.
Are there any quantities that are not invariable, and what would their importance be?

Of central importance is the relation of topological features with functionality.
The optimal network covering leads to the partitioning of the network into boxes.
Do these boxes carry a message other than nodes proximity? For example, the boxes could be used as
an alternative definition for separated communities, and fractal methods could
be used as a novel method for community detection in networks~\cite{newman4,clauset,jim1,jim2,vicsek2}.

The networks that we have presented are all static, with no temporal component,
and time evolution has been ignored in all our discussions above.
Clearly, biological networks, the WWW, and other networks have grown (and continue to grow)
from some earlier simpler state to their present fractal form. Has fractality always been
there or has it emerged as an intermediate stage obeying certain evolutionary drive forces?
Is fractality a stable condition or growing networks will eventually fall into a non-fractal form?

Finally, we want to know what is the inherent reason behind fractality. Of course, we have already
described how hub-hub anti-correlations can give rise to fractal networks. However, can this be
directly related to some underlying mechanism, so that we gain some information on the process?
In general, in Biology we already have some idea on the advantages of adopting a fractal structure.
Still, the question remains: why fractality exists in certain networks and not in others?
Why both fractal and non-fractal networks are needed? It seems that we will be able to
increase our knowledge for the network evolutionary mechanisms through fractality studies.

In conclusion, a deeper understanding of the self-similarity, fractality and transfractality of complex networks
will help us analyze and better understand many fundamental properties of real-world networks.

\section{APPENDIX: The Box Covering Algorithms}

The estimation of the fractal dimension and the self-similar features in networks have become standard properties
in the study of real-world systems. For this reason, in the last three years many box covering algorithms have been proposed~\cite{sornette, song3}.
This section presents four of the main algorithms, along with a brief discussion on the advantages and disadvantages that they offer.

Recalling the original definition of box covering by Hausdorff \cite{bunde-havlin,feder,Peitgen}, for
a given network $G$ and box size $\ell_B$, a box is a set of
nodes where all distances $\ell_{ij}$ between any two nodes i and j in the box
are smaller than $\ell_B$. The minimum number of boxes required to cover
the entire network $G$ is denoted by $N_B$. 
For $\ell_B = 1$, each box encloses only 1 node and therefore, $N_B$ is equal to the size of the
network $N$. On the other hand, $N_B=1$ for $\ell_B \ge \ell_B^\mathrm{max}$, where $\ell_B^\mathrm{max}$ is the diameter of the network plus one.

The ultimate goal of a box-covering algorithm is to find the \emph{minimum} number of boxes $N_B(\ell_B)$ for any $\ell_B$. It has been shown that this problem belongs to the family of NP-hard problems \cite{Garey}, which means that the solution cannot be achieved in polynomial time.
In other words, for a relatively large network size, there is no algorithm that can provide an exact solution in a reasonably short amount of time.
This limitation requires treating the box covering problem with approximations, using for example optimization algorithms.

\subsection{The greedy coloring algorithm}

The box-covering problem can be mapped into another NP-hard problem \cite{Garey}: the graph coloring problem.

\begin{figure}
\centerline{
 {\resizebox{6cm}{!} { \includegraphics{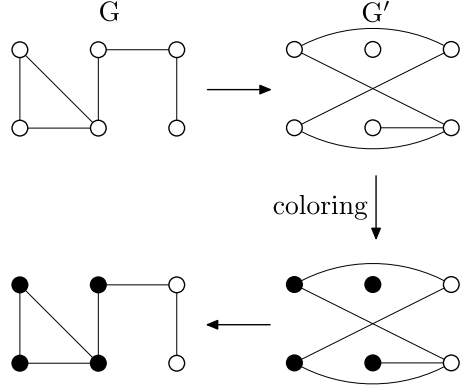}}}
} \caption {Illustration of the solution for the network covering problem
via mapping to the graph coloring problem. Starting
from $G$ (upper left panel) we construct the dual network $G'$ (upper right panel) for a given box size (here $\ell_B=3$), where two nodes are connected if they are
at a distance $\ell\geq\ell_B$. We use a greedy algorithm for vertex coloring in $G'$,
which is then used to determine the box covering in $G$, as shown in the plot.}
\label{aux}
\end{figure}

An algorithm that approximates well the optimal solution of this problem was presented in \cite{song3}. For an arbitrary value of
$\ell_B$, first construct a dual network $G'$, in which
two nodes are connected if the distance between
them in $G$ (the original network) is greater or equal than $\ell_B$.
Fig.~\ref{aux} shows an example of a network $G$
which yields such a dual network $G'$ for $\ell_B=3$ (upper row
of Fig.~\ref{aux}).

Vertex coloring is a well-known procedure, where labels (or colors) are assigned to
each vertex of a network, so that no edge connects two
identically colored vertices. It is clear that such a coloring in
$G'$ gives rise to a natural box covering in the original
network $G$, in the sense that vertices of the same color will necessarily form a
box since the distance between them must be less than $\ell_B$.
Accordingly, the minimum number of boxes $N_B(G)$ is equal to the
minimum required number of colors (or the chromatic number) in the dual network $G'$,
$\chi(G')$.

In simpler terms, (a) if the distance between two nodes in $G$ is greater than $\ell_B$
these two neighbors cannot belong in the same box. According to the construction
of $G'$, these two nodes will be connected in $G'$ and thus they cannot have
the same color. Since they have a different color they will not belong in the
same box in $G$. (b) On the contrary,
if the distance between two nodes in $G$ is less than $\ell_B$
it is possible that these nodes belong in the same box. In $G'$ these
two nodes will not be connected and it is allowed for these two nodes to carry the same color,
i.e. they may belong to the same box in $G$, (whether these nodes will actually be connected
depends on the exact implementation of the coloring algorithm).

The algorithm that follows both constructs the dual network $G'$ and assigns
the proper node colors for all $\ell_B$ values in one go.
For this implementation a two-dimensional matrix $c_{i\ell}$ of size $N\times \ell_B^{\rm max}$ is needed, whose values represent the color of node $i$ for a given box size $\ell=\ell_B$.
\begin{enumerate}
  \item Assign a unique id from 1 to N to all network nodes, without assigning any colors yet.
  \item For all $\ell_B$ values, assign a color value 0 to the node with id=1, i.e. $c_{1\ell}=0$.
  \item Set the id value $i=2$. Repeat the following until $i=N$.
  \begin{enumerate}
    \item Calculate the distance $\ell_{ij}$  from $i$ to all the
    nodes in the network with id $j$ less than $i$.
    \item Set $\ell_B=1$
    \item Select one of the unused colors $c_{j\ell_{ij}}$ from all nodes
    $j<i$ for which $\ell_{ij}\geq\ell_B$. This is the color $c_{i\ell_B}$ of node $i$ for the
    given $\ell_B$ value.
    \item Increase $\ell_B$ by one and repeat (c) until $\ell_B=\ell_B^\mathrm{max}$.
    \item Increase $i$ by 1.
  \end{enumerate}
\end{enumerate}

The results of the greedy algorithm may depend on the original coloring
sequence. The quality of this algorithm was investigated by randomly reshuffling the coloring sequence and applying the greedy algorithm several times and in different models~\cite{song3}. The result was that the probability distribution of the number of boxes $N_B$ (for all box sizes $\ell_B$) is a narrow Gaussian distribution, which indicates that almost any implementation of the algorithm yields a solution close to the optimal.

Strictly speaking, the calculation of the fractal dimension $d_B$ through
the relation $N_B\sim \ell_B^{-d_B}$ is valid only for the minimum possible
value of $N_B$, for any given $\ell_B$ value, so any box covering algorithm must aim to
find this minimum $N_B$.
Although there is no rule to determine when this minimum value has been
actually reached (since this would require an exact solution of the
NP-hard coloring problem) it has been shown~\cite{Cormen} that the greedy coloring
algorithm can, in many cases, identify a coloring sequence which yields
the optimal solution.

\subsection{Burning algorithms}

This section presents three box covering algorithms based on more traditional
breadth-first search algorithm.

\begin{figure}
\centerline{
{\resizebox{6cm}{!} { \includegraphics{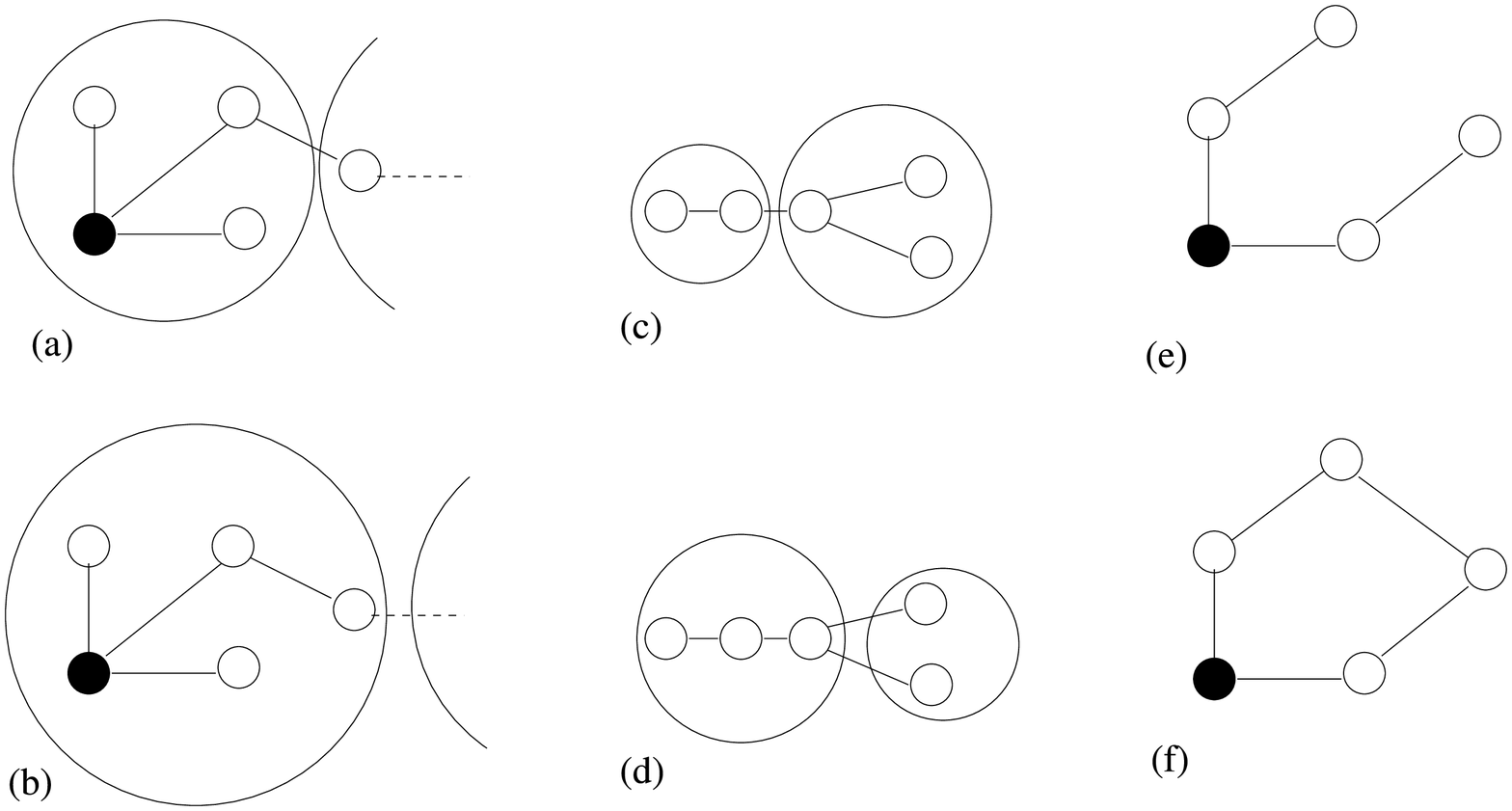}}}
} \caption {Our definitions for a box that is (a) non-compact for $\ell_B=3$, i.e. could include more nodes,
(b) compact, (c) connected, and (d) disconnected (the nodes in the right
box are not connected in the box). (e) For this box, the values $\ell_B=5$ and $r_B=2$
verify the relation $\ell_B=2r_B+1$. (f) One of the pathological cases
where this relation is not valid, since $\ell_B=3$ and $r_B=2$.} \label{definitions}
\end{figure}

A box is defined as \emph{compact} when it includes the maximum
possible number of nodes, i.e. when there do not exist any other
network nodes that could be included in this box. A \emph{connected} box means that
any node in the box can be reached from any other node in this box,
without having to leave this box. Equivalently, a \emph{disconnected} box
denotes a box where certain nodes can be reached by other nodes in the box
only by visiting nodes outside this box. For a demonstration of these definitions
see Fig.~\ref{definitions}.

\subsection{Burning with the diameter $\ell_B$, and the Compact-Box-Burning (CBB) algorithm}

The basic idea of the CBB algorithm for the generation of a box is to start from a given box center and then expand
the box so that it includes the maximum possible number of nodes,
satisfying at the same time the maximum distance between nodes in the box $\ell_B$.
The CBB algorithm is as follows (see Fig.~\ref{FIGdraw_CBB}):

\begin{figure}
\centerline{
   {\resizebox{9cm}{!} { \includegraphics{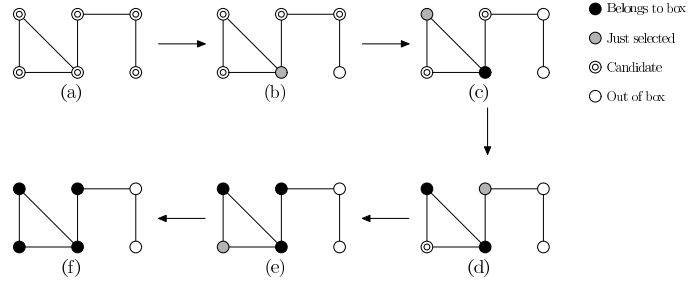}}}
} \caption {Illustration of the CBB algorithm for $\ell_B=3$. (a) Initially, all nodes are candidates for the box.
(b) A random node is chosen, and nodes at a distance further than $\ell_B$ from this
node are no longer candidates. (c) The node chosen in (b) becomes part of the box and another candidate
node is chosen. The above process is then repeated until the box is complete.} \label{FIGdraw_CBB}
\end{figure}

  \begin{enumerate}
  
  \item Initially, mark all nodes as uncovered.
  
   \item Construct the set $C$ of all yet uncovered nodes.

   \item Choose a random node $p$ from the set of uncovered nodes $C$ and
   remove it from $C$.

   \item Remove from $C$ all nodes $i$ whose distance from $p$ is $\ell_{pi}\geq\ell_B$,
   since by definition they will not belong in the same box.

   \item Repeat steps (3) and (4) until the candidate set is empty.
   
   \item Repeat from step (2) until all the network has been covered.
  \end{enumerate}


\subsection{Random Box Burning}

In 2006, J. S. Kim \emph{et al.} presented a simple algorithm for the calculation of fractal dimension in networks~\cite{kim1,kim2, kim00}:
  \begin{enumerate}
   \item Pick a randomly chosen node in the network as a seed of the box.

   \item Search using breath-first search algorithm until distance $l_B$ from the seed. Assign all newly burned nodes to the new box. If no new node is found, discard and start from (1) again.
   
   \item Repeat (1) and (2) until all nodes have a box assigned.
   \end{enumerate}

This Random Box Burning algorithm has the advantage of being a fast and simple method. However, at the same time
there is no inherent optimization employed during the network coverage. Thus, this simple Monte-Carlo method is almost
certain that will yield a solution far from the optimal and one needs to implement many different realizations
and only retain the smallest number of boxes found out of all these realizations. 

\subsection{Burning with the radius $r_B$, and the Maximum-Excluded-Mass-Burning (MEMB) algorithm}

A box of size $\ell_B$ includes nodes where the distance between any pair of nodes is less than $\ell_B$.
It is possible, though, to grow a box from a given central node, so that all nodes in the box are within distance less than a given box radius $r_B$
(the maximum distance from a central node). This way, one can still recover the same fractal properties of a network.
For the original definition of the box, $\ell_B$ corresponds to the box diameter
(maximum distance between any two nodes in the box) plus one.
Thus, $\ell_B$ and $r_B$ are connected through the simple relation $\ell_B = 2 r_B+1$. In general this relation is exact for loopless configurations,
but in general there may exist cases where this equation is not exact (Fig.~\ref{definitions}).

The MEMB algorithm always yields the optimal solution for non scale-free homogeneous networks, since the choice of the central node is not important.
However, in inhomogeneous networks with wide-tailed degree distribution, such as scale-free networks, this algorithm fails to achieve an optimal
solution because of the presence of hubs.

\begin{figure}
\centerline{
   {\resizebox{6cm}{!} { \includegraphics{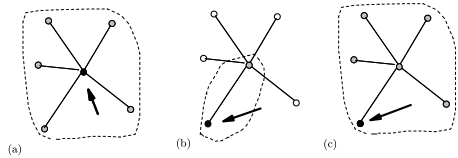}}}
} \caption {Burning with the radius $r_B$ from (a) a hub node or (b) a non-hub node results
in very different network coverage. In (a) we need just one box of $r_B=1$
while in (b) 5 boxes are needed to cover the same network. This is
an intrinsic problem when burning with the radius. (c) Burning
with the maximum distance $\ell_B$ (in this case $\ell_B=2r_B+1=3$) we avoid
this situation, since independently of the starting point we would
still obtain $N_B=1$.} \label{burning}
\end{figure}

The MEMB, as a difference from the Random Box Burning and the CBB, attempts to locate some \emph{optimal central}
nodes which act as the burning origins for the boxes. It contains as a special case the choice of the hubs as centers of the boxes,
but it also allows for low-degree nodes to be burning centers, which sometimes is convenient for finding a solution closer to the optimal.

\begin{figure}
\centerline{
   {\resizebox{9cm}{!} { \includegraphics{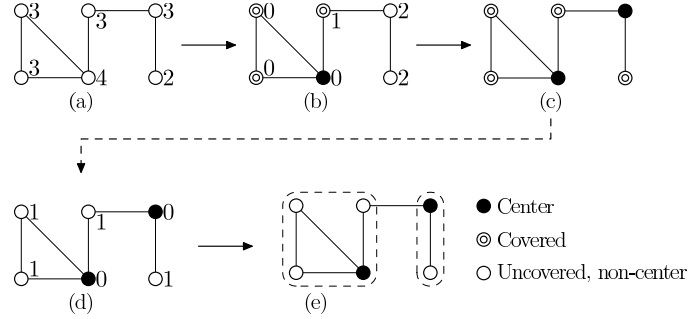}}}
} \caption {Illustration of the MEMB algorithm for $r_B=1$. {\it Upper row: Calculation of the box centers}
(a) We calculate the excluded mass for each node. (b) The node with maximum mass becomes a center
and the excluded masses are recalculated. (c) A new center is chosen. Now, the entire network is covered
with these two centers. {\it Bottom row: Calculation of the boxes} (d) Each box includes initially only the center.
Starting from the centers we calculate the distance of each network node to the closest center.
(e) We assign each node to its nearest box.} \label{FIGdraw_MEMB}
\end{figure}

In the following algorithm we use the basic idea of box optimization, in which each box covers the maximum possible number of nodes. For a given burning radius $r_B$, we define the \emph{excluded mass} of a node as the number of uncovered nodes within a chemical distance less than $r_B$. First, calculate the excluded mass for all the uncovered nodes. Then, seek to cover the network with boxes of maximum excluded mass. The details of this algorithm are as follows (see Fig.~\ref{FIGdraw_MEMB}):

\begin{enumerate}
  \item Initially, all nodes are marked as uncovered and non-centers.
  \item For all non-center nodes (including the already covered nodes)
  calculate the excluded mass, and select the node $p$ with
  the maximum excluded mass as the next center.
  \item Mark all the nodes with chemical distance less than $r_B$ from $p$ as covered.
  \item Repeat steps (2) and (3) until all nodes are either covered or centers.
\end{enumerate}

Notice that the excluded mass
has to be updated in each step because it is possible that it has been modified during this step.
A box center can also be an already covered node, since it may lead to a larger box mass.
After the above procedure, the number of selected centers coincides with the number
of boxes $N_B$ that completely cover the network.
However, the non-center nodes have not yet been assigned to a given box.
This is performed in the next step:

\begin{enumerate}
  \item Give a unique box id to every center node.
  \item For all nodes calculate the ``central distance", which is the chemical distance to its
  nearest center. The central distance has to be less than $r_B$, and the center
  identification algorithm above guarantees that there will always exist such a center.
  Obviously, all center nodes have a central distance equal to 0.
  \item Sort the non-center nodes in a list according to increasing central distance.
  \item For each non-center node $i$, at least one of its neighbors has a
  central distance less than its own. Assign to $i$ the same id with this
  neighbor. If there exist several such neighbors, randomly select an id
  from these neighbors. Remove $i$ from the list.
  \item Repeat step (4) according to the sequence from the list in step (3) for all non-center nodes.
\end{enumerate}


\subsection{Comparison between algorithms}

The choice of the algorithm to be used for a problem depends on the details of the problem itself. If connected boxes are a requirement,
MEMB is the most appropriate algorithm; but if one is only interested in obtaining the fractal dimension of a network, the greedy-coloring or the random box burning are more suitable since they are the fastest algorithms.

\begin{figure}
\centerline{
   {\resizebox{8cm}{!} { \includegraphics{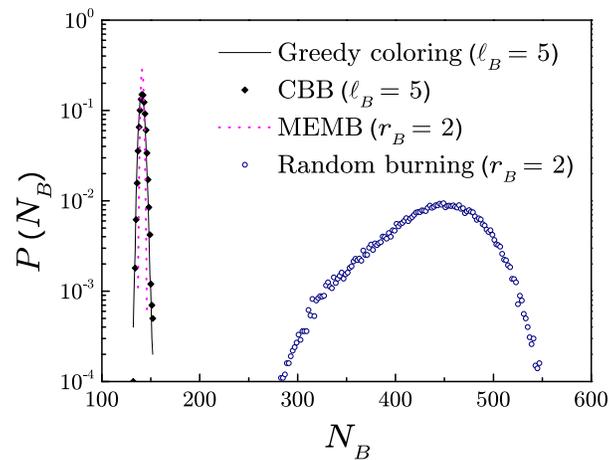}} }
} \caption {Comparison of the distribution of $N_B$ for $10^4$ realizations of the four network covering methods presented in
this paper. Notice that three of these methods yield very similar results with narrow distributions
and comparable minimum values, while the random burning algorithm fails to reach a value close to this minimum
(and yields a broad distribution).
 } \label{comp_all}
\end{figure}

As explained previously, any algorithm should intend to find the optimal solution, that is, find the minimum number of boxes that cover the network. Fig.~\ref{comp_all} shows the performance of each algorithm. The greedy-coloring, the CBB and MEMB algorithms exhibit a narrow distribution of the number of boxes, showing evidence that they cover the network with a number of boxes that is close to the optimal solution. Instead, the Random Box Burning returns a wider distribution and its average is far above the average of the other algorithms. Because of the great ease and speed with which this technique can be implemented, it would be useful to show that the average number of covering boxes is overestimated by a fixed proportionality constant. In that case, despite the error, the predicted number of boxes would still yield the correct scaling and fractal dimension. 

\end{document}